\documentclass[aps,prb,twocolumn,superscriptaddress]{revtex4}
\usepackage{ae}
\usepackage[T1]{fontenc}
\usepackage[ansinew]{inputenc}
\usepackage{amsmath}
\usepackage{amssymb}
\usepackage{graphicx}
\usepackage{color}

\usepackage[colorlinks]{hyperref}
\usepackage{epstopdf}

\def\el{{\mbox{\boldmath{$\ell$}}}}

\usepackage{soul,xcolor}
\setstcolor{red}                   

\def\sig{{\mbox{\boldmath{$\sigma$}}}}

\begin{document}

\date{\today}

\title{Spin geometric-phases in hopping magnetoconductance }

\author{O. Entin-Wohlman}
\email{entin@tau.ac.il}
\affiliation{Raymond and Beverly Sackler School of Physics and Astronomy, Tel Aviv University, Tel Aviv 69978, Israel}
\author{A. Aharony}
\affiliation{Raymond and Beverly Sackler School of Physics and Astronomy, Tel Aviv University, Tel Aviv 69978, Israel}

\begin{abstract}
We identify  theoretically the geometric phases of the electrons' spinor that can be detected in measurements of charge and spin transport through Aharonov-Bohm interferometers threaded by a magnetic flux $\Phi$ (in units of the flux quantum) where   Rashba spin-orbit and Zeeman interactions are active.
We show that the combined effect of these  two interactions produces a $\sin(\Phi)$  [in addition to the usual $\cos(\Phi)$] dependence of the magnetoconductance, whose amplitude is proportional to the Zeeman field.
Therefore, although the magnetoconductance is an even function of the magnetic field, it  is not a periodic function of it, and the widely-used concept of a phase shift
in the Aharonov-Bohm oscillations, as indicated in previous work, is not applicable. We find the directions of the  spin-polarizations  in the system, and show that in general (even without the Zeeman term) the spin currents are not conserved, implying the generation of magnetization in the terminals attached to the interferometer.

\end{abstract}

\maketitle

\section{Introduction}
\label{intro}
The phase factor induced on the spin wave-functions of  electrons moving in  an electric field   is generally attributed to the Aharonov-Casher effect, \cite{AC}  the electromagnetic dual of the Aharonov-Bohm effect. \cite{AB} The physical origin of this phenomenon is the
spin-orbit interaction, that gives rise to a momentum-dependent magnetic field perpendicular to both the propagation direction of the electron waves and the electric field  which creates it.
Attempts to monitor the Aharonov-Casher phase factor utilize various  designs of spin interferometers in low-dimensional electronic structures. Of particular significance in this context    is the
 spin-orbit interaction generated by the
Bychkov-Rashba mechanism,  \cite{Bychkov} which can be controlled via  asymmetric gate electrodes. \cite{Takayanagi}
Indeed,   measuring the resistance of an array of
mesoscopic semiconductor rings placed in
an asymmetric quantum well
whose asymmetry was tuned by a gate voltage,
enabled the demonstration of
the Aharonov-Casher interference.  \cite{Bergsten}  The value of the spin-orbit interaction parameter which was deduced from the interference patterns was  in agreement with the one derived from   Shubnikov-de Haas analyses (measured in the presence of orbital magnetic fields).

The Aharonov-Casher phase comprises two contributions. \cite{Qian} Besides the standard,
 ``dynamic phase", related to the precession of the spins around a constant magnetic-field direction, there appears also a geometric (topological)  phase: the Berry phase.  \cite{Berry,Vanderbilt} It is accumulated when the direction of the  spin-orbit-induced  magnetic field follows   adiabatically a closed trajectory in space and 
 is proportional to the solid angle subtended by the direction of that field. 
Under non-adiabatic conditions the spin precession axis is no longer parallel to the spin-orbit-induced magnetic field. Then the acquired phse is termed   the Aharonov-Anandan phase \cite{Anandan} (which generalizes the Berry phase).

 The geometric phase  on  the spin wave-function  and the way it affects physical observables have been the subject of intense  theoretical and experimental research in recent years. This phase  was predicted to lead to topological persistent spin-currents in closed loops even in the absence of  electromagnetic fluxes  \cite{Loss,Oh}  or in rings coupled spin-symmetrically to external reservoirs. \cite{Ellner}
As the Aharonov-Casher phase  modifies the transmission of electrons through multiply-connected mesoscopic structures, 
\cite{Avishai} it is natural to explore its effects on  Aharonov-Bohm oscillations measured on  such devices.
In the absence of the spin-orbit interaction, and neglecting the Zeeman field on the spin magnetic moment, the magnetoconductance of Aharonov-Bohm interferometers is periodic in the magnetic flux $\Phi$ (measured in units of the flux quantum) which is due to an external  magnetic field normal to the plane of the interferometer.
It was shown \cite{Meir} that
the spin-orbit interaction modifies the magnetic-flux $\Phi$   dependence of the
spectrum of electrons in a one-dimensional ring into a $\Phi\pm\gamma$ dependence for the  two relevant spin
directions. Here,  $\gamma$ is the Aharonov-Casher phase that results from the spin-orbit interaction.
A spin-interference device,  in which the phase difference between the spin wave-functions traveling in the clockwise and anti-clockwise directions can be measured,  was indeed  proposed. \cite{Nitta} The idea was elaborated upon theoretically, \cite{Peeters,Frustaglia,Capozza} and measurements were carried out   in a GaAs two-dimensional hole system with  a strong spin-orbit coupling,  \cite{Yau}
and in ring structures fabricated from HgTe/HgCdTe quantum wells. \cite{Konig}
The Aharonov-Casher phase was also detected in measurements of
interference patterns of the magnetoconductance  of electrons traveling coherently in opposite directions around arrays of InAlAs/InGaAs mesoscopic  semiconductor rings. \cite{Nagasawa} 

While Aharonov-Bohm  conductance oscillations allow for the experimental detection of the Aharonov-Casher phase,  attempts to  separate the
 dynamic and topological aspects of this phase  rely on additional information,  extracted either from simulations \cite{Yau}
or from comparisons with theoretical expressions  \cite{Qian,Frustaglia} which express the geometric and  dynamic phases in terms of the Aharonov-Casher  phase {\em  calculated for a circular  ring}.
For such rings, the Aharonov-Casher phase indeed determines both the dynamic and the geometric Aharonov-Anandan phases. In contrast,  we find that for  general configurations there exist no such simple relations, and it may not be possible to extract the dynamic and geometric phases from the  Aharonov-Bohm interference patterns.

This paper is aimed to identify the geometric phases of the electrons' spinor that can be detected in measurements of  charge and spin transport through
Aharonov-Bohm interferometers in which both the Rashba spin-orbit  and  Zeeman interactions are active (the latter is due to a magnetic field normal to the plane of the  interferometer).
In particular, we consider the possible appearance of a phase shift in the (periodic) dependence of the magnetoconductance on the Aharonov-Bohm flux $\Phi$, and the phases which determine the directions of the spin magnetizations in the two reservoirs connected to the interferometer. The phase shift  of $\Phi$ is especially intriguing:
perhaps the first  to propose
that  the spin-orbit and Zeeman interactions join together to shift the Aharonov-Bohm oscillations for each spin direction by the Berry phase
were Aronov and Lyanda-Geller. \cite{Lyanda}
Operating within the adiabatic limit where both interaction energies are much larger than the angular kinetic energy, they predicted phase shifts in the Aharonov-Bohm oscillations of the transmission of each spin direction.  Aronov and Lyanda-Geller related these phase shifts to  the Berry phase that depends on the Zeeman energy. A similar conclusion appears in Ref. \onlinecite{Peeters}, where this deviation  is attributed to the Aharonov-Casher phase.
Reference \onlinecite{Lyanda} has been criticised in the literature,  because it used a non-hermitian Hamiltonian,  \cite{Morpurgo}
and because it was claimed \cite{Yi}
that it had not accounted properly for the effects of the time-reversal symmetry breaking by the Zeeman field. We do not share this second argument: our results below, though obtained in an entirely different regime,   indicate that Grosso Modo the transmission given in  Ref. \onlinecite{Lyanda} has the correct form, namely, the (longitudinal) conductance  is an even function of the magnetic field, as required by the Onsager relations, though it is not periodic in it.  However, we find that in general the deviation away from periodicity  is not related to the Berry phase. 

In contrast to the theoretical works surveyed above, which adopt  a continuous description for the Hamiltonian of the electrons on a circular ring (with the exception of Ref. \onlinecite{Avishai}), we consider transport in the hopping regime. In other words, the kinetic energy of the electrons when on the  interferometer loop is modelled in a tight-binding picture. This necessitates the detailed knowledge of the tunneling matrix elements between localized states
when the tunneling electrons are subjected to  spin-orbit interactions and  Zeeman and orbital magnetic fields, which we derive in Appendix \ref{amplitude}.
Calculating the charge (particle)  current through an arbitrarily-shaped single-channel triangular interferometer coupled to two non-polarized electronic reservoirs,
we find that quite generally, the {\em interference-related} part of the electrical conductance
{\em in the absence of the Zeeman interaction}
 is proportional to $\cos(\Phi)\cos(\gamma)$,
as indeed has been predicted.  \cite{Avishai,Meir}
When the Zeeman interaction is included  the pattern of the Aharonov-Bohm
oscillations is modified: there appears a $\sin(\Phi)$ term in the interference-induced transmission, whose amplitude is
proportional to the Zeeman field and which vanishes without the spin-orbit interaction. The coefficient  of this term
depends on details of the triangular structure. Although this field-dependent amplitude breaks the periodicity of the conductance in the magnetic field, thus eliminating the concept of a phase shift,  the entire conductance still
 obeys the Onsager relations. In this respect we agree with the final result of Ref. \onlinecite{Lyanda}. However, there is no obvious relation between this amplitude and the angles that determine the spin-magnetization directions. Therefore, as opposed to the finding of Ref. \onlinecite{Lyanda}, in general measuring the interference pattern of  Aharonov-Bohm oscillations yields no information on the Berry phase.
As Qian and Su \cite{Qian} show, the limit of a regular polygon with $N\rightarrow\infty$ edges approaches the perfect circle, and it is possible that for that  geometry one recovers the Aronov-Lyanda-Geller \cite{Lyanda} predictions. However, we do not get such results even for an equilateral triangle, and therefore we suspect that the circle is very special, and it is not possible to deduce the Berry phase from transport measurements on a general interferometer.

The Berry phase is usually associated with the direction of the electron's spin polarization once it tunnels around the closed loop. It turns out that for a circular loop the spin-orbit interaction indeed induces a non-zero spin polarization, that makes a tilt-angle $\chi$ with the $\hat{\bf z}-$axis that is perpendicular to the plane of the loop. 
For a circular ring and in the absence of the Zeeman interaction, the angle $\chi$ determines both the Aharonov-Anandan phase $\pm \pi [1-\cos(\chi)]$ (with the sign related to the spin state along the polarization axis) and the dynamic phase $\pm \tan (\chi)\sin(\chi)$. \cite{Qian} The former is one half of the solid angle subtended by the rotating polarization. \cite{Lyanda}
Our calculations for an arbitrary triangular loop show that in the absence of the Zeeman interaction,  the details of the spin polarization components in the left ($L$) and right ($R$) terminals depend on two {\it different} angles  $\chi^{}_L$ and $\chi^{}_R$. These two angles have no simple relation between them, and there is no relation between them and the Berry phase. However,  the magnitudes of the spin polarizations in the two terminals are
proportional to $\sin(\Phi)\sin(\gamma)$, and thus they are periodic and odd in the magnetic field. In the presence of the Zeeman field $B$, there appears, in addition  to the $\sin(\Phi)$ dependence, an additional  term in the spin polarizations,  proportional to $B\cos\Phi$ (which is again odd in the magnetic field). This term breaks the periodicity in $\Phi$ and excludes the use of the concept of a phase shift.

The paper is organized as follows. Section \ref{calc} outlines the calculation of the transmission through the interferometer. It begins with  the definitions of the particle and  spin currents  (Sec. \ref{PMrate}), presents the model Hamiltonian (Sec. \ref{Hamil}), and then continues with the calculation of the transmission matrix  (in spin space) which determines the currents (Sec. \ref{rate}). The details of this calculation are given in Appendix \ref{GF}.
In Sec.  \ref{trans}  we analyze the interference-induced terms in  the particles' current, that is, in the charge conductance, and in the spin magnetization rates. In the first part, Sec. \ref{trnoB},  we consider the effects of the Rashba interaction alone, and point out which geometric phases can be extracted from interference data. In the second part, Sec. \ref{trB}, we analyze the joint effects of the spin-orbit and the Zeeman interactions. The details of this calculation are given in Appendix \ref{triangle}. Section \ref{con} contains a summary of the results and the ensuing conclusions.

\section{Transmission }
\label{calc}

\subsection{Particle and magnetization rates}
\label{PMrate}

Our model system is the standard one: it comprises two leads coupled through a central region. Electrons moving through the central region are subjected to  the Rashba spin-orbit interaction and to a Zeeman field.
We also include  the effect of an Aharonov-Bohm flux $\Phi$ (in units of the flux quantum) in the resulting expressions.
For unpolarized leads  the Fermi distribution function, e.g., in the left lead, is
\begin{align}
f^{}_{L}(\epsilon^{}_{k})=[e^{\beta(\epsilon ^{}_{k}-\mu^{}_{L})}+1]^{-1}_{}\ ,
\label{Fermi dis}
\end{align}
where
$\beta=(k^{}_{\rm B}T)^{-1}$ is the inverse temperature, $\epsilon^{}_{k}$ is the single-electron energy in the left lead,  and $\mu_{L}$ is the chemical potential there. The Fermi distribution in the right reservoir, $f^{}_{R}(\epsilon^{}_{p})$, is defined similarly.

The rate of change of the particles number in the left lead, $I^{L}$, i.e. the particle current,  is
\begin{align}
I^{L}_{}=\frac{d}{dt}\sum_{\sigma}\sum_{\bf k}\langle c^{\dagger}_{{\bf k}\sigma}c^{}_{{\bf k}\sigma}\rangle\ ,
\label{IL}
\end{align}
where $c^{\dagger}_{{\bf k}\sigma}$ ($c^{}_{{\bf k}\sigma}$) creates (annihilates) a particle  with momentum ${\bf k}$ and spin $\sigma$ (at an arbitrary quantization-axis at this stage) in the left lead; the angular brackets indicate a quantum average. Similar definitions pertain for the right lead, with ${\bf k}$ replaced by ${\bf p}$.
 (We use units in which $\hbar=1$.) Since our system is at steady state, charge is conserved, i.e., $I^{L}=-I^{R}$.
 This is not the case for the magnetization rates. The rate of change of the magnetization   in the left lead, $\dot{\bf M}^{L}_{}$,  which can be interpreted as the  spin current in that lead, is  \cite{Tokura}
\begin{align}
\frac{d}{dt}{\bf M}^{L}_{}\equiv\dot{\bf M}^{L}_{}=\frac{d}{dt}\sum_{\sigma, \sigma'}\sum_{\bf k}\langle c^{\dagger}_{{\bf k}\sigma}c^{}_{{\bf k}\sigma'}\rangle[\sig]^{}_{\sigma\sigma'}\
\label{SC}
\end{align}
(in dimensionless units).  Here $\sig$ is the vector of the Pauli matrices.
The magnetization rate in the right lead, $\dot{\bf M}^{R}_{}$, is defined similarly. In addition to $\dot{\bf M}^{L}$ and $\dot{\bf M}^{R}$, the magnetization is also changed in the central region; hence the spin currents in the leads are not necessarily conserved.

Both the particle current and the magnetization rate in the left lead  are determined  by the rate $R^{L}_{\sigma\sigma'}$
\begin{align}
R^{L}_{\sigma\sigma '}=
\frac{d}{dt}\sum_{\bf k}\langle c^{\dagger}_{{\bf k}\sigma}c^{}_{{\bf k}\sigma'}\rangle\ .
\label{R}
\end{align}
This quantity is calculated below within the Keldysh technique. 

\subsection{The model Hamiltonian}
\label{Hamil}

\begin{figure}[htp]
\includegraphics[width=8cm]{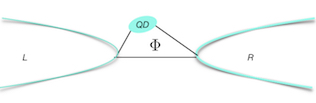}
\caption{(Color online.)  Illustration of our model. A triangular Aharonov-Bohm interferometer  is  threaded by a magnetic flux $\Phi$ (in units of the flux quantum). The interferometer is coupled  to two reservoirs. One of its arms  carries a quantum dot. A  Zeeman field normal to the  plane of the triangle affects both the dot and the arms of the loop. An electric field normal to the plane of the triangle generates a Rashba spin-orbit interaction on the links forming the triangle.}
\label{ring1}
\end{figure}

The model system is illustrated in Fig. \ref{ring1}: an interferometer formed of three straight  segments is coupled to two electronic reservoirs. The reservoirs (when decoupled) are not spin-polarized; they are described by free electron gases, with the Hamiltonian
\begin{align}
{\cal H}^{}_{\rm leads}=\sum_{{\bf k},\sigma}\epsilon^{}_{k}c^{\dagger}_{{\bf k}\sigma}c^{}_{{\bf k}\sigma}+
\sum_{{\bf p},\sigma}\epsilon^{}_{p}c^{\dagger}_{{\bf p}\sigma}c^{}_{{\bf p}\sigma}\ .
\end{align}
One arm of the interferometer connects the two leads directly, and the other carries a quantum dot, represented by a localized level of energy
\begin{align}
 \epsilon^{}_{\sigma}\equiv \epsilon^{}_{0}-\sigma B\ ,
\end{align}
which includes the Zeeman energy,  $B$,  on the dot. It is assumed that the Zeeman field $B$ (in units of energy) is along the $\hat{\bf z}-$direction, normal to the plane of the triangle. The Hamiltonian of the quantum dot is
\begin{align}
{\cal H}^{}_{\rm dot}=\sum_{\sigma}\epsilon^{}_{\sigma} d^{\dagger}_{\sigma}d^{}_{\sigma}\ ,
\end{align}
where $d^{\dagger}_{\sigma} $ ($d^{}_{\sigma}$) creates (annihilates) an electron in the state $|\sigma\rangle$ on the dot. The tunneling Hamiltonian that joins together all these components, is
\begin{align}
{\cal H}^{}_{\rm tun}&=
\sum_{\sigma,\sigma'}(\sum_{\bf k}e^{i\phi^{}_{L}}[V^{}_{{\bf k}d}]^{}_{\sigma\sigma'}c^{\dagger}_{{\bf k}\sigma}d^{}_{\sigma'}
\nonumber\\
&+\sum_{\bf p}e^{-i\phi^{}_{R}}[V^{}_{{\bf p}d}]^{}_{\sigma\sigma'}c^{\dagger}_{{\bf p}\sigma}d^{}_{\sigma'})\nonumber\\
&+
\sum_{\sigma,\sigma'}\sum_{{\bf k},{\bf p}}[V^{}_{{\bf k}{\bf p}}]^{}_{\sigma\sigma'}c^{\dagger}_{{\bf k}\sigma}c^{}_{{\bf p}\sigma'}+{\rm H.c.}\ .
\label{Htun}
\end{align}
The first  two terms on the right hand-side of Eq. (\ref{Htun}) represent the tunneling between the dot and the leads; $\phi_{L(R)}$ is the partial Aharonov-Bohm phase  acquired from the orbital magnetic field (in units of the flux quantum $\Phi_{0}$) along the link between the left lead and the dot (from the dot to the right lead), such that
\begin{align}
\phi^{}_{L}+\phi^{}_{R}=\Phi
\end{align}
is the total flux  penetrating the loop.
The last term
describes the direct tunneling between the two leads.  The three  amplitudes $V$,  2$\times$2 matrices in spin space,  include the effect of the spin-orbit  and  Zeeman interactions on the tunneling  (see Appendix \ref{amplitude}).
They are discussed in detail below. At this stage it is enough to assume that each of them allows for spin-flip processes of the electron while it tunnels on the edges of the triangle. It is assumed further that
 the tunneling amplitudes are calculated on the Fermi surface, and therefore are independent of ${\bf k}$ and ${\bf p}$, except for the dependence on the direction of the link; this is encoded in the dependence on the spin indices (see Appendix \ref{amplitude}). Hence
 \begin{align}
V^{}_{\bf kp}\rightarrow V^{}_{LR}\ ,\ \ V^{}_{{\bf k}d}\rightarrow V^{}_{Ld}\ ,\ \ V^{}_{{\bf p}d}\rightarrow V^{}_{Rd}\ .
\label{vap}
\end{align}

\subsection{The particle and spin currents}
\label{rate}

Introducing the definition of the lesser Green's function
\begin{align}
[G^{<}_{{\bf k}d}(t,t')]^{}_{\sigma\sigma'}=i\langle d^{\dagger}_{\sigma'}(t')c^{}_{{\bf k}\sigma}(t)\rangle\ ,
\end{align}
(with similar definitions for the subscripts of all other Green's functions that appear below), we can write the rate (\ref{R}) in the left lead in the form
\begin{align}
&R^{L}_{\sigma\sigma'}=\int\frac{d\omega}{2\pi}\sum_{\sigma^{}_{1}}\{\sum_{{\bf k},{\bf p}}
(
 [V^{\ast}_{{\bf k}{\bf p}}]^{}_{\sigma\sigma^{}_{1}}
[G^{<}_{{\bf k}{\bf p}}(\omega)]^{}_{\sigma'\sigma^{}_{1}}\nonumber\\
&-
 [V^{}_{{\bf k}{\bf p}}]^{}_{\sigma'\sigma^{}_{1}}[G^{<}_{{\bf p}{\bf k}}(\omega)]^{}_{\sigma^{}_{1}\sigma})
+\sum_{\bf k}( e^{-i\phi^{}_{L}}[V^{\ast}_{{\bf k}d}]^{}_{\sigma\sigma^{}_{1}}[G^{<}_{{\bf k}d}(\omega)]^{}_{\sigma'\sigma^{}_{1}}\nonumber\\
&-e^{i\phi^{}_{L}}[V^{}_{{\bf k}d}]^{}_{\sigma'\sigma^{}_{1}}[G^{<}_{d{\bf k}}(\omega)]^{}_{\sigma^{}_{1}\sigma})\}\ .
\label{rate1}
\end{align}
As the system is stationary,  the Green's functions depend only on the time difference. Hence, it is convenient to use the Fourier transforms (for all Green's functions)
\begin{align}
[G^{<}_{{\bf k}d}(t,t')]^{}_{\sigma\sigma'}=\int\frac{d\omega}{2\pi}e^{-i\omega(t-t')}
[G^{<}_{{\bf k}d}(\omega)]^{}_{\sigma\sigma'}
\ .
\end{align}
Exploiting  relations (\ref{vap}), and using matrix notations (in spin space), Eq. (\ref{rate1}) for the rate takes the form
\begin{align}
[R^{L}_{}]^{\rm t}_{}&=\int \frac{d\omega}{2\pi}[G^{<}_{LR}(\omega)V^{\dagger}_{LR}-V^{}_{LR}G^{<}_{RL}(\omega)\nonumber\\
&+e^{-i\phi^{}_{L}}G^{<}_{Ld}(\omega)V^{\dagger}_{Ld}-e^{i\phi^{}_{L}}V^{}_{Ld}G^{<}_{dL}(\omega)]
\ ,
\label{ratem}
\end{align}
where the superscript t on $[R^{L}_{}]^{\rm t}_{}$ indicates the transposed matrix, and
where
\begin{align}
&G^{}_{dL}(\omega)=\sum_{\bf k}G^{}_{d{\bf k}}(\omega)\ ,\ \
G^{}_{Ld}(\omega)=\sum_{\bf k}G^{}_{{\bf k}d}(\omega)\ ,\nonumber\\
&G^{}_{LR}(\omega)=\sum_{{\bf k},{\bf p}}G^{}_{{\bf k}{\bf p}}(\omega)\ ,\ \  G^{}_{RL}(\omega)=\sum_{{\bf k},{\bf p}}G^{}_{{\bf p}{\bf k}}(\omega)\ , 
\label{defG}
\end{align}
are matrices in spin space. These Green's functions are derived  in terms of the tunneling amplitudes and the Green's functions of the decoupled system (which are denoted by lowercase letters);   see Appendix    \ref{GF} for the detailed calculation.
We show there [Eq. (\ref{ratef})]
that to  lowest order in the tunneling amplitudes the rate  (\ref{ratem}) is
\begin{align}
[R^{L}_{}]^{\rm t}_{}&=4\pi^{2}_{}J^{2}{\cal N}^{}_{L}{\cal N}^{}_{R}\int \frac{d\omega}{2\pi}[f^{}_{R}(\omega)-f^{}_{L}(\omega)]{\cal T}^{}_{L}(\omega)\ ,
\label{RLL}
\end{align}
with the transmission   matrix (in spin space),
 \begin{align}
{\cal T}^{}_{L}(\omega)&={\cal A}^{}_{LR}(\omega){\cal A}^{}_{RL}(\omega)
\ .
\label{tau}
\end{align}
Here, $J$ is the energy that scales the tunneling amplitudes $V$ (see Appendix \ref{amplitude}), ${\cal N}^{}_{L(R)}$ is the density of states at the Fermi energy in the left (right) lead,
and
\begin{align}
{\cal A}^{}_{LR}(\omega)&={\cal A}^{\dagger}_{RL}(\omega)=\frac{1}{J}\Big (V^{}_{LR}+e^{i\Phi}V^{}_{Ld}g^{}_{d}(\omega)V^{\dagger}_{Rd}\Big )\ ,
\label{A}
\end{align}
where $g^{}_{d}(\omega)$
 is the Green's function of the decoupled dot, Eq. (\ref{gd}). The transmission matrix ${\cal T}^{}_{R}(\omega)$ is obtained from Eq. (\ref{tau}) by replacing $R\leftrightarrow L$, and $\Phi\leftrightarrow -\Phi$.

The particle current  in the left lead is thus [see Eqs. (\ref{IL}) and (\ref{R})]
\begin{align}
I^{L}_{}=2\pi J^{2}_{}{\cal N}^{}_{L}{\cal N}^{}_{R}\int d\omega[f^{}_{R}(\omega)-f^{}_{L}(\omega)]{\rm Tr}\{{\cal T}^{}_{L}(\omega)\}\ .
\label{IL1}
\end{align}
The particle current in the right lead, $I^{R}$,  is derived from Eqs. (\ref{A}) and (\ref{IL1}) by replacing $R\leftrightarrow L$ and $\Phi\leftrightarrow -\Phi$,  which results in the transmission ${\cal T}^{}_{R}(\omega)$ multiplied by
$[f^{}_{L}(\omega)-f^{}_{R}(\omega)]$.
As ${\rm Tr}\{{\cal T}^{}_{L}(\omega)\}={\rm Tr}\{{\cal T}^{}_{R}(\omega)\}$,
the particle currents
obey $I^{L}=-I^{R}$, as required for charge conservation. This is  not the case with the spin currents.
The spin current in the left lead [Eqs. (\ref{SC}) and (\ref{RLL})] is
\begin{align}
\dot{\bf M}^{L}_{}=
2\pi J^{2}_{}{\cal N}^{}_{L}{\cal N}^{}_{R}\int d\omega[f^{}_{R}(\omega)-f^{}_{L}(\omega)]{\rm Tr}\{{\cal T}^{}_{L}(\omega)\sig\}\
\label{ML1}
\end{align}
and the one in the right lead is derived from Eq. (\ref{ML1}) by replacing $R\leftrightarrow L$ {and $\Phi\leftrightarrow -\Phi$.
 Since in general
${\cal A}_{LR}$ [Eq. (\ref{A})] does not commute with ${\cal A}_{RL}$, the spin currents  in the two leads are not necessarily conserved (see Sec. \ref{trnoB}).


\section{Interference contributions to the transmission}
\label{trans}
As seen from Eqs. (\ref{tau}) and (\ref{A}), the transmission matrix pertaining to the left lead  ${\cal T}^{}_{L}(\omega)$ comprises two parts. The first, ${\cal T}^{0}_{L}(\omega)$,
is independent of the Aharonov-Bohm flux,
\begin{align}
{\cal T}^{0}_{L}(\omega)=\frac{1}{J^{2}}\Big (V^{}_{LR}V^{\dagger}_{LR}+V^{}_{Ld}g^{}_{d}(\omega)V^{\dagger}_{Rd}
V^{}_{Rd}g^{}_{d}(\omega)V^{\dagger}_{Ld}\Big )\ .
\label{T0}
\end{align}
It describes the transmission through the direct connection between the reservoirs, and the transmission through the arm carrying the quantum dot.
The second part of the  left-side transmission matrix (\ref{tau}), ${\cal T}^{\rm in}_{L}(\omega,\Phi)$, is the interference contribution,
\begin{align}
{\cal T}^{\rm in}_{L}(\omega,\Phi)&=[
e^{-i\Phi}V^{}_{LR}V^{}_{Rd}g^{}_{d}(\omega)V^{\dagger}_{Ld}+{\rm H.c.}]/J^{2}
\ ,
\label{TLR}
\end{align}
and that corresponding to the interference contribution of the right-side transmission ${\cal T}^{\rm }_{R}$ reads
\begin{align}
{\cal T}^{\rm in}_{R}(\omega,\Phi)&=[
e^{i\Phi}V^{}_{RL}V^{}_{Ld}g^{}_{d}(\omega)V^{\dagger}_{Rd}+{\rm H.c.}]/J^{2}
\ .
\label{TRL}
\end{align}
One observes that ${\rm Tr}\{{\cal T}^{\rm in}_{L}(\omega,\Phi)\}=
{\rm Tr}\{{\cal T}^{\rm in}_{R}(\omega,\Phi)\}$; however, generally
${\rm Tr}\{{\cal T}^{\rm in}_{L}(\omega,\Phi)\sig\}\neq
{\rm Tr}\{{\cal T}^{\rm in}_{R}(\omega,\Phi)\sig\}$, indicating  that while the particle current is conserved, generally, the spin current in the two leads  is not [see Eqs. (\ref{IL1}) and (\ref{ML1})}].
It is interesting to compare the electronic paths that constitute
${\cal T}^{\rm in}_{L}(\omega,\Phi)$ and ${\cal T}^{\rm in}_{R}(\omega,\Phi)$: Eq. (\ref{TLR}) comprises paths that
traverse the interferometer from the $L$ point and backwards (see Fig. \ref{ring1}), while ${\cal T}^{\rm in}_{R}(\omega,\Phi)$
contains the paths that start and end at the point $R$.


\subsection{Interference transport: effects of the Rashba interaction}
\label{trnoB}
Here we discuss the transmissions and the  accompanying geometric phases in the case where the electrons on the interferometer are subjected solely to the Rashba interaction and the orbital Aharonov-Bohm flux. The tunneling amplitude pertaining to this situation is derived in Appendix \ref{amplitude} [see Eq. (\ref{prop1})]; it is found that in the absence of the Zeeman field the tunneling amplitudes are proportional to unitary matrices,
\begin{align}
V^{}_{Ld}/J&= \cos(k^{}_{\rm so}d^{}_{Ld})+i\sin(k^{}_{\rm so}d^{}_{Ld})
\hat{\bf e}^{}_{Ld}\cdot\sig\ ,\nonumber\\
V^{}_{Rd}/J&= \cos(k^{}_{\rm so}d^{}_{Rd})+i\sin(k^{}_{\rm so}d^{}_{Rd})
\hat{\bf e}^{}_{Rd}\cdot\sig\ ,\nonumber\\
V^{}_{LR}/J&= \cos(k^{}_{\rm so}d^{}_{LR})+i\sin(k^{}_{\rm so}d^{}_{LR})
\hat{\bf e}^{}_{LR}\cdot\sig
\ ,
\label{e}
\end{align}
where $J$ sets the energy scale of the tunneling (assumed for simplicity to be the same on all edges\cite{comm}).
Here, $k_{\rm so}$ is the strength of the Rashba interaction (in momentum units), $d_{ij}$ is the length of the corresponding edge of the triangle, with $i,j=L,d,R$ (see Fig. \ref{ring1}), and $\hat{\bf e}_{ij}$ is a unit vector in the direction of the effective magnetic field induced by the
Rashba interaction on an electron tunneling through the edge $d_{ij}$,
\begin{align}
\hat{\bf e}^{}_{ij}=\hat{\bf n}\times\hat{\bf d}^{}_{ij}\ .
\label{ee}
\end{align}
Here,  $\hat{\bf d}^{}_{ij}$ is a unit vector in the direction of the bond from $i$ to $j$, and $\hat{\bf n}$ is the direction of the electric field creating the Rashba interaction. In our model $\hat{\bf n}$ is normal to the plane of the triangle, and is chosen to lie along the $\hat{\bf z}$ axis; hence all three vectors $\hat{\bf e}_{ij}$ lie in the plane of the triangle, i.e. in the $x-y-$plane.


Given the explicit forms of the $V$'s in Eq. (\ref{e}), we can now  discuss the relations between the magnetization rates in the two leads, $\dot{\bf M}^{L,R}_{}$.
Starting from Eq. (\ref{TRL}), we find
\begin{widetext}
\begin{align}
&{\rm Tr}\{{\cal T}^{\rm in}_{R}(\omega,\Phi)\sig\cdot\hat{\el}\}
={\rm Tr}\{
e^{-i\Phi}V^{}_{Rd}g^{}_{d}(\omega)V^{\dagger}_{Ld}V^{}_{LR}\sig\cdot\hat{\el}+{\rm H.c.}\}/J^{2}\ ,
\label{TRLsiga}
\end{align}
where $\hat{\el}$ is an arbitrary unit vector. For $\hat{\el}=\hat{\bf e}^{}_{LR}$, $\sig\cdot\hat{\el}$ commutes with $V^{}_{LR}$, and therefore
\begin{align}
&{\rm Tr}\{{\cal T}^{\rm in}_{R}(\omega,\Phi)\sig\cdot\hat{\bf e}^{}_{LR}\}
={\rm Tr}\{
e^{-i\Phi}V^{}_{Rd}g^{}_{d}(\omega)V^{\dagger}_{Ld}\sig\cdot\hat{\bf e}^{}_{LR} V^{}_{LR}+{\rm H.c.}\}/J^{2}\nonumber\\
&={\rm Tr}\{
e^{-i\Phi}V^{}_{LR}V^{}_{Rd}g^{}_{d}(\omega)V^{\dagger}_{Ld}\sig\cdot\hat{\bf e}^{}_{LR}+{\rm H.c.}\}/J^{2}
={\rm Tr}\{{\cal T}^{\rm in}_{L}(\omega,\Phi)\sig\cdot\hat{\bf e}^{}_{LR}\}\ .
\label{TRLsig}
\end{align}
\end{widetext}
Substituting this relation in Eq. (\ref{ML1}) yields
\begin{align}
\dot{\bf M}^{L}\cdot\hat{\bf e}^{}_{LR}&=-\dot{\bf M}^{R}\cdot\hat{\bf e}^{}_{LR}\ ,
 \label{MMM1}
  \end{align}
  which means that the  current of the spin component perpendicular to the $LR$ bond in the interferometer plane is conserved. Note that in the absence of the Zeeman interaction ${\cal T}^{0}_{L}(\omega)$ does not contribute to the magnetization rates.

The situation is more complicated for the two other spin components, perpendicular to $\hat{\bf e}^{}_{LR}$. These components
do not commute (or anti-commute) with $V^{}_{LR}$, and therefore they are not conserved. Indeed, a direct calculation using the commutator $[V^{}_{LR},\sig]$ shows that $\dot{\bf M}^{L}+\dot{\bf M}^{R}$ is non-zero  in the plane  perpendicular to $\hat{\bf e}^{}_{LR}$. This sum represents a magnetization which is injected into the two reservoirs.
As we show in Appendix \ref{triangle}, there are no simple relations between these components in the two reservoirs (except for the equilateral triangle).  Within our low-order expansion in the tunneling amplitudes we neglect the effects of these polarizations on the electronic distribution functions in the reservoirs, Eq. (\ref{Fermi dis}). However, there will be some buildup of spin polarizations in the terminals near the interferometer, which might be detectable experimentally.

We next relate the particle and spin currents to the various angles and phases.
As in the absence of the Zeeman field $g_{d}(\omega)=[\omega-\epsilon^{}_{0}]^{-1}$ is just a number, it is clear that
$V^{}_{LR}V^{}_{Rd}V^{}_{dL}/J^{3}$ (note that $V^{\dagger}_{Ld}=V^{}_{dL}$), which corresponds to a tunneling path
from the left lead, around the triangle, back to the left lead,
 is also a unitary matrix, and has the general form
\begin{align}
V^{}_{LR}V^{}_{Rd}V^{}_{dL}/J^{3}=C^{}_{L}+i{\bf c}^{}_{L}\cdot\sig\ ,
\label{p1}
\end{align}
where  $C^{}_{L}$ is a real scalar and ${\bf c}^{}_{L}$ is a real vector, that has components in the plane of the triangle,  ${\bf c}^{}_{L,\perp}$, as well as a normal component, $c_{L,z}$,
\begin{align}
{\bf c}^{}_{L}={\bf c}^{}_{L,\perp}+\hat{\bf z} c^{}_{L,z}\ .
\label{ckatan}
\end{align}
Explicit expressions for these coefficients for an arbitrary triangle are given in  Eq. (\ref{noZ}).
As seen from Eqs. (\ref{SC}), (\ref{RLL}),  and (\ref{TLR}), the magnetization rate in
the left lead (or the spin current associated with that lead) is along the vector ${\bf c}^{}_{L}$ [see also Eq. (\ref{ML2}) below]. Therefore, we consider this vector as an effective magnetic field, induced by the spin-orbit interaction which generates this magnetization.
The appearance of the  $z-$component indicates that the effective magnetic field
(which on each edge is in the plane of the triangle) attains a $z-$component when the amplitude along a path formed of several edges is followed.
Interestingly, this component arises from the non-commutability of the tunneling matrices on the three triangle edges.
 \cite{Avishai}
 We denote by $\chi^{}_{L}$ the tilt-angle between this spin-orbit-induced  magnetic field and the $\hat{\bf z}-$axis,
\begin{align}
\tan (\chi^{}_{L})=|{\bf c}^{}_{L,\perp}|/c^{}_{L,z}\ .
\label{chi1}
\end{align}
 For a perfectly circular geometry, this angle is related to the Berry phase, or more generally to the Aharonov-Anandan  phase, see e.g., Refs. {\onlinecite{Qian}, \onlinecite{Vanderbilt},  and  \onlinecite{Frustaglia}. As already mentioned, this is not true for the arbitrary triangular interferometer.

A priori, the right-to-right path requires the product
\begin{align}
V^{}_{RL}V^{}_{Ld}V^{}_{dR}/J^{3}=C^{}_{R}+i{\bf c}^{}_{R}\cdot\sig\ ,
\label{p2}
\end{align}
 and consequently
\begin{align}
\tan (\chi^{}_{R})=|{\bf c}^{}_{R,\perp}|/c^{}_{R,z}\ .
\label{chi2}
\end{align}
On a physical basis, one expects that
\begin{align}
C^{}_{L}=C^{}_{R}\equiv C\ ,
\label{CLR}
 \end{align}
 as is indeed the case, since ${\rm Tr}\{V^{}_{LR}V^{}_{Rd}V^{}_{dL}\}={\rm Tr}\{V^{}_{RL}V^{}_{Ld}V^{}_{dR}\}$ is a real number [see also Eq. (\ref{noZ})].
Explicitly, Eqs.  (\ref{p1}) and
(\ref{chi1}) can be written as
\begin{align}
&V^{}_{LR}V^{}_{Rd}V^{}_{dL}/J^{3}=C\nonumber\\
&+i\frac{c^{}_{L,z}}{\cos(\chi^{}_{L})}\left[\begin{array}{cc}\cos(\chi^{}_{L})&\sin(\chi^{}_{L})e^{-i\varphi^{}_{L}} \\
\sin(\chi^{}_{L})e^{i\varphi^{}_{L}}&
-\cos(\chi^{}_{L})\end{array}\right ]\ ,
\label{p22}
\end{align}
where we denote ${\bf c}^{}_{L,\perp}=|{\bf c}^{}_{L,\perp}|[\cos(\varphi^{}_{L})\hat{\bf x}+\sin(\varphi^{}_{L})\hat{\bf y}]$. The spinor eigen vectors of the matrix in Eq. (\ref{p22}), whose corresponding eigen values are $\pm 1$,
are well-known,
\begin{align}
|+\rangle =\left [\begin{array}{c}\cos (\chi^{}_{L}/2) \\ \sin(\chi^{}_{L}/2) e^{i\varphi^{}_{L}}\end{array}\right ]\ ,\
|-\rangle =\left [\begin{array}{c}-\sin (\chi^{}_{L}/2) \\ \cos(\chi^{}_{L}/2) e^{i\varphi^{}_{L}}\end{array}\right ]\ .
\label{eigv}
\end{align}
It hence follows that
\begin{align}
V^{}_{LR}V^{}_{Rd}V^{}_{dL}/J^{3}=e^{i\gamma^{}_L}|+\rangle\langle+|+e^{-i\gamma^{}_L}|-\rangle\langle-|\ ,
\label{p3}
\end{align}
where
\begin{align}
\tan (\gamma^{}_{L})=c^{}_{L,z}/[C\cos(\chi^{}_{L})]=|{\bf c}^{}_{L}|/C
\end{align}
is the Aharonov-Casher phase. \cite{AC}
Due to the unitarity of the matrices in conjunction  with Eq. (\ref{CLR}),
$C_{}^{2}+|{\bf c}^{}_{L}|^{2}=1=C_{}^{2}+|{\bf c}^{}_{R}|^{2}$,
the Aharonov-Casher phase $\gamma^{}_{L}$ acquired by the  left-to-left path and the  one accumulated on the  right-to-right path, $\gamma_{R}$,  are identical,
\begin{align}
\gamma^{}_{L}=\gamma^{}_{R}\equiv \gamma\ .
\label{gamLR}
\end{align}
Therefore,
\begin{align}
{\cal T}^{\rm in}_{L}(\omega,\Phi)&=\frac{J}{\omega-\epsilon^{}_{0}}\Big (
e^{-i\Phi}[e^{i\gamma}|+\rangle\langle+|+e^{-i\gamma}|-\rangle\langle-|]\nonumber\\
&+
e^{i\Phi}[e^{-i\gamma}|+\rangle\langle+|+e^{i\gamma}|-\rangle\langle-|]
\Big )\ ,
\end{align}
and hence
\begin{align}
{\rm Tr}\{{\cal T}^{\rm in}_{L}(\omega,\Phi)\}&=
{\rm Tr}\{{\cal T}^{\rm in}_{R}(\omega,\Phi)\}\nonumber\\
&=
\frac{4J}{\omega-\epsilon^{}_{0}}\cos(\Phi)\cos(\gamma)\ .
\label{tr}
\end{align}
The  interference part of the particle currents [Eq. (\ref{IL1})] is thus
\begin{align}
I^{L,{\rm in}}_{}=-I^{R,{\rm in}}_{}&=8\pi J^{3}_{}{\cal N}^{}_{L}{\cal N}^{}_{R}\cos(\Phi)\cos(\gamma)\nonumber\\
&\times\int d\omega\frac{[f^{}_{R}(\omega)-f^{}_{L}(\omega)]}{\omega -\epsilon^{}_{0}}\ .
\label{IL2}
\end{align}
There is no phase shift in $\Phi$, but measuring the amplitude of the periodic and symmetric Aharonov-Bohm  oscillations in the magnetoconductance gives information about the Aharonov-Casher phase of the interferometer, via the coefficient $\cos(\gamma)$. This observation was made a long time ago, \cite{Meir} and was recently examined in great detail. \cite{Avishai}

The
spin current associated with  the left lead is given by Eq. (\ref{ML1}), with
\begin{align}
{\rm Tr}\{{\cal T}^{\rm in}_{L}(\omega,\Phi)\sig\}=&\frac{4J\sin(\Phi)\sin(\gamma)}{\omega-\epsilon^{}_{0}}\nonumber\\
&\times[\hat{\bf z}\cos(\chi^{}_{L})+\hat{\bf c}^{}_{L,\perp}\sin(\chi^{}_{L})]\ ,    
\label{TLRS}
\end{align}
which is parallel to the vector ${\bf c}^{}_L$. Hence,
\begin{align}
&\dot{\bf M}^{L}_{}=
8\pi J^{3}_{}{\cal N}^{}_{L}{\cal N}^{}_{R}\sin(\Phi)\sin(\gamma)\nonumber\\
&\times\int d\omega
\frac{f^{}_{R}(\omega)-f^{}_{L}(\omega)}{\omega-\epsilon^{}_{0}}
[\hat{\bf z}\cos(\chi^{}_{L})+\hat{\bf c}^{}_{L,\perp}\sin(\chi^{}_{L})]\ .
\label{ML2}
\end{align}
Similar to the calculation presented above, the magnetization rate in the right lead is given by
 \begin{align}
&\dot{\bf M}^{R}_{}=
8\pi J^{3}_{}{\cal N}^{}_{L}{\cal N}^{}_{R}\sin(\Phi)\sin(\gamma)\nonumber\\
&\times\int d\omega
\frac{f^{}_{R}(\omega)-f^{}_{L}(\omega)}{\omega-\epsilon^{}_{0}}
[\hat{\bf z}\cos(\chi^{}_{R})+\hat{\bf c}^{}_{R,\perp}\sin(\chi^{}_{R})]\ ,
\label{ML2a}
\end{align}
which is along ${\bf c}^{}_R$.
From Eq. (\ref{MMM1})  it then follows that
${\bf c}^{}_L\cdot\hat{\bf e}^{}_{LR}= -{\bf c}^{}_R\cdot\hat{\bf e}^{}_{LR}$, but there exists no obvious relation between the transverse components in the two reservoirs, and thus no relation between the angles $\chi^{}_L$ and $\chi^{}_R$.

Several conclusions  can be drawn from the expressions for the interference-induced parts of the currents. The conspicuous one is that  measuring the interference parts of the currents in an  arbitrarily-shaped Aharonov-Bohm interferometer in the absence of a Zeeman field gives unambiguously only the Aharonov-Casher \cite{AC} geometric phase $\gamma$. Unlike the case of the circular loop, \cite{Qian,Frustaglia} it is not clear if and how measurements of  the tilting angles $\chi^{}_L$ and $\chi^{}_R$ can yield information about the Aharonov-Anandan (or the Berry) or the dynamic phases.

\vspace{.5cm}


\subsection{Interference transport: joint effects of spin-orbit and  Zeeman interactions}
\label{trB}

When the tunneling electrons are subjected to the Zeeman interaction alone then [see Appendix \ref{triangle}, in particular Eqs. (\ref{noSO}) and (\ref{dotZ})]
\begin{align}
&V^{}_{LR}V^{}_{Rd}g^{}_{d}(\omega)V^{}_{dL}=V^{}_{RL}V^{}_{Ld}g^{}_{d}(\omega)V^{}_{dR}=S^{}_{1}+S^{}_{2}\sigma^{}_{z}\ ,
\end{align}
where $S_{1}$ and $S^{}_{2}$  are real expressions comprising $\cosh(\kappa d^{}_{ij})$,
$\sinh(\kappa d^{}_{ij})$,  and $B/(\omega-\epsilon^{}_0)$.
Here,
$d^{}_{ij}$ is the length of the bond $ij$, see Fig. \ref{ring1},
$g_{d}(\omega)$ is the Green's function of the decoupled dot [given in Eqs. (\ref{gd}) and (\ref{gdZ})], and $\kappa=m^{\ast}Ba$ is derived in Appendix \ref{amplitude}
($a$ is
the localization radius in the barrier \cite{Shahbazyan}  and $m^{\ast}$ is the effective mass).
It follows that the transmissions ${\cal T}^{\rm in}_{L}$  and
${\cal T}^{\rm in}_{R}$
[Eqs. (\ref{TLR}) and (\ref{TRL}), respectively]
are identical, and are proportional to
$\cos(\Phi)(S^{}_{1}+S^{}_{2}\sigma^{}_{z})$. Obviously,  the interference-induced parts of the charge conductance and the magnetization rates have solely the $\cos(\Phi)$ dependence; the magnetization rates are  only along $\hat{\bf z}$, the direction of the Zeeman field. Both the charge and the spin currents are conserved.
One also notes that the oscillatory dependence on the lengths $d_{ij}$ of the bonds, resulting from the Rashba interaction,   is totally lost in this case
(it is replaced by a combination of hyperbolic functions).

The presence of both the spin-orbit and the Zeeman  interactions modifies  the results reported in Sec. \ref{trnoB}. Since the pertaining expressions are relatively cumbersome, and in any event depend heavily on geometric details,  they are relegated to Appendix \ref{triangle}. Here we summarize the qualitative results.
We begin with the interference contribution to the charge conductance. Let us first ignore the Zeeman interaction on the decoupled dot and keep this interaction  only in the tunneling amplitudes. Then from Eq. (\ref{triple})
we find that there are real contributions to $V^{}_{LR}V^{}_{Rd}V^{}_{dL}$, which are second order in
$\overline{B}=m^{\ast}Ba/k^{}_{2}$, with  \cite{comim}     $k^{}_{2}=\sqrt{k^{2}_{\rm so}-(m^{\ast}Ba)^{2}}$, see Appendix \ref{amplitude}. Consequently, these terms are multiplied by $\cos(\Phi)$. These additional contributions to the charge conductance   modify  the  coefficient of $\cos(\Phi)$ in Eq. (\ref{IL2}) but do not change it in any essential way; they do change the period of the oscillations with the lengths $d^{}_{ij}$. \cite{t0}

The effect of  the terms  in  $V^{}_{LR}V^{}_{Rd}V^{}_{dL}$ which are  first order in $\overline{B}$ is very different. Since they lead to one of our main results, we reproduce them here. From Eq. (\ref{triple}), the contributions to the scalar part $C$ are
\begin{widetext}
\begin{align}
&i\overline{k}^{2}_{\rm so}\overline{B}
\sin(k^{}_{2}d^{}_{LR})\sin(k^{}_{2}d^{}_{Rd})\sin(k^{}_{2}d^{}_{dL})
[\hat{\bf e}^{}_{LR}\times\hat{\bf e}^{}_{Rd}\cdot\hat{\bf z}
+\hat{\bf e}^{}_{dL}\times\hat{\bf e}^{}_{LR}\cdot\hat{\bf z}+
\hat{\bf e}^{}_{Rd}\times\hat{\bf e}^{}_{dL}\cdot\hat{\bf z}]
\ .
\label{in}
\end{align}
As seen, the contributions of the terms in
Eq. (\ref{in}) to the charge conductance are multiplied by $\overline{B}\sin(\Phi)$, making the  expression even in the magnetic field, but not periodic. 
For example,  confining ourselves to an equilateral triangle,
we find that Eq. (\ref{IL2}) changes to
\begin{align}
I^{L,{\rm in}}_{}=-I^{R,{\rm in}}_{}&=8\pi J^{3}{\cal N}^{}_{L}{\cal N}^{}_{R}\int d\omega\frac{f^{}_{R}(\omega)-f^{}_{L}(\omega)}{\omega-\epsilon^{}_{0}}
\nonumber\\
&\times\Big (\Big [\cos^{3}_{}(k^{}_{2}d)
+3\cos(k^{}_{2}d)\sin^{2}(k^{}_{2}d)[\overline{B}^{2}+\overline{k}^{2}_{\rm so}/2]\Big ]\cos(\Phi)
-3\sqrt{3}\sin^{3}(k^{}_{2}d)\overline{k}^{2}_{\rm so}\overline{B}\sin(\Phi)\Big )\ ,
\end{align}
\end{widetext}
where $\overline{k}^{}_{\rm so}=k^{}_{\rm so}/k^{}_{2}$. 
 The authors of Refs. \onlinecite{Peeters} and \onlinecite{Lyanda}  expressed the conductance as a sum of two contributions (from the two spin directions), each of which is a periodic function in  the Aharonov-Bohm flux with its own phase shift that depends on the Zeeman field.
In Ref. \onlinecite{Peeters} that phase shift is related to the Aharonov-Casher phase of the interferometer, while that of Ref. \onlinecite{Lyanda} is claimed to be the Berry phase. As seen, our result  is generally not related  neither to the tilting angles $\chi^{}_L$ or $\chi^{}_R$ discussed in Sec. \ref{trnoB}, nor to the Aharonov-Casher phase.  In this respect our result is also different from the one of Ref. \onlinecite{Ady}, derived for  a system subjected to a Zeeman field perpendicular to the loop and another, time-dependent magnetic field which rotates in the plane of the loop: the expression for the conductivity  presented there contains only  $\cos (\Phi)$ terms.

Next we discuss what modifications  the Zeeman interaction on the dot may introduce. As  shown in Appendix \ref{triangle},  apart from real terms whose contribution to the charge conductance is multiplied by $\cos(\Phi)$, there appear also terms of the type presented in Eq. (\ref{in}), with $\overline{B}$ replaced by $B/(\omega-\epsilon^{}_{0})$. This implies that the Zeeman interaction on the dot alone,  in conjunction with the spin-orbit interaction in the tunneling amplitudes, suffices to induce the $\sin(\Phi)$ dependence of the  Aharonov-Bohm oscillations alluded to above. Strictly speaking, one expects that $\overline{B}>B/(\omega-\epsilon^{}_{0})$  when
 the dot is  far from resonance (as assumed in our model).

The effect of the Zeeman interaction on the magnetization rates is discussed in the second part of Appendix \ref{triangle}. Since the magnetization rate is expected to be odd in the magnetic field, it is not surprising that the terms which are linear in the Zeeman field contribute only terms proportional to $\cos(\Phi)$ to the magnetization rates.

\section{conclusions and discussion}
\label{con}

We have analyzed  the effects of the Rashba spin-orbit and Zeeman interactions  on the flux dependence of the charge and spin transport through   Aharonov-Bohm interferometers.
 We find that the  Zeeman interaction, which  is often   ignored in the calculations,  causes crucial qualitative changes in the results, which should be observable.

Most of the earlier literature concentrated on circular loops,   found relations among the geometric Aharonov-Casher, Aharonov-Anandan and dynamic phases, and considered possibilities to extract these phases from experiments. We find that the circular configuration is probably unique, and  simple relations seem not to exist for other shapes of the loop. To demonstrate this point we calculated the charge and spin currents through a triangular loop, to lowest order in the  tunneling amplitudes.

The spin-orbit interaction is known to generate an effective spin-dependent vector potential, so that a spin which moves around a loop accumulates the Aharonov-Casher phase $\gamma$. When this interaction is added to the Aharonov-Bohm interferometer (but without the Zeeman field), we find that the leading interference contribution to the total magnetoconductance of the interferometer (coupled to unpolarized leads) is proportional to $\cos\Phi\cos\gamma$, and the only way to extract $\gamma$ is to study the amplitude of the $\cos\Phi$ terms. No other phase (e.g., Aharonov-Anandan or Berry) can be extracted from such measurements. 

The spin-orbit interaction, even without the Zeeman interaction,  also generates interesting spin currents in the leads, whose amplitudes are all proportional to $\sin\Phi\sin\gamma$. Thus, measuring  these currents again yields only the phase $\gamma$. While the current of the spin component in the plane of the triangle, perpendicular to the edge $LR$ which connects the terminals directly, is conserved, the spin-orbit interaction on the interferometer generates a growing magnetization of the other two spin components in the two terminals. In general, these magnetizations  have different tilt angles $\chi^{}_L$ and $\chi^{}_R$ with the axis perpendicular to the interferometer plane and different projections on the interferometer plane. These directions depend on the detailed structure of the triangular loop. We found no simple relation between the tilt angles and the Aharonov-Casher phase $\gamma$, or with any other ``standard" phases. In fact, Vanderbilt \cite{Vanderbilt}
calculated the Berry phase for an equilateral triangle and for a regular  $N-$edge polygon (Exercise 3.1.2 in Ref. \onlinecite{Vanderbilt}), and found that it can be characterized by a single tilt angle $\chi$. His result reduces in the $N\rightarrow \infty$ limit to the circular loop result $\pm\pi(1-\cos\chi)$. However, the expression for the Berry phase pertaining to a finite regular polygon is more complex, and clearly all these phases depend on the details of  structure of the interferometer loop.

Adding a Zeeman field perpendicular to the interferometer plane breaks the unitarity of the tunneling amplitudes on the interferometer edges, and therefore also on the amplitude of a spinor after it goes around the loop. For a non-unitary matrix, one can no longer use the concept of the Aharonov-Casher phase.
Indeed, we find that  the amplitude of the $\cos\Phi$ function in the total magnetoconductance is modified by a term which depends on the Zeeman field.  To obtain the pure spin-orbit Aharonov-Casher phase one has  to extrapolate this amplitude to zero field. Furthermore, the combination of the spin-orbit interaction and the Zeeman field generates a new term in the magnetoconductance, which is proportional to $B\sin\Phi$. If one could treat the Zeeman field $B$ and the Aharonov-Bohm flux $\Phi$ as two independent parameters  of the problem, which can be fixed separately, then one could interpret this term as generating a phase shift in $\Phi$, proportional to $B$. However, in practice both the Aharonov-Bohm flux and the Zeeman interactions arise due to the same external magnetic field, and $\Phi\propto B$.  The resulting magnetoconductance is no longer a periodic function, but it remains an even function of the field. Since the phase $\Phi$ results only from the normal component of the field, it may be interesting to consider rotations of the field away from this normal direction. For symmetry reasons, we expect the magnetoconductance of any interferometer  to  have the aperiodic but even form $a^{}_1+a^{}_2\cos\Phi+a^{}_3 B\sin\Phi$, with corrections of order $B^2$ in each coefficient $a^{}_i$, with a similar odd analog for the spin currents.


\begin{acknowledgments}
This research was  partially supported by the Israel Science Foundation (ISF), by the infrastructure program of Israel Ministry of Science and Technology under contract 3-11173, and  by the Pazy Foundation. We acknowledge the hospitality of the PCS at IBS, Daejeon, Korea, where part of this work was done, under IBS funding number (IBS-R024-D1).
\end{acknowledgments}


\appendix


\section{The Green's functions' calculation }
\label{GF}

The Green's functions are derived from the corresponding Dyson's equations,  in terms of the tunneling amplitudes and the Green's functions of the decoupled system (which are denoted by lowercase letters). The Green's functions of the decoupled  reservoirs are independent of the spin indices,
\begin{align}
g^{}_{L(R)}(\omega)=\sum_{k(p)}g^{}_{k(p)}(\omega)\ ,
\end{align}
and that of the decoupled dot is diagonal in spin space
\begin{align}
g^{}_{d}(\omega)=[\omega-\epsilon^{}_{0}+\sigma^{}_{z}B]^{-1}
\label{gd}
\end{align}
(it is assumed that the decoupled dot is empty).
The lesser superscript is omitted, as the Dyson equations are valid for the three Keldysh Green's functions, lesser, retarded and advanced;  these are found by using  Langreth rules \cite{Langreth} for the analytic continuation  on the time  contour. \cite{Jauho}

The Dyson equations for $G^{}_{Ld}$ and $G^{}_{dL}$ [see Eqs. (\ref{defG})], which involve the Green's function
$G^{}_{LL}(\omega)=\sum_{{\bf k},{\bf k}^{}_{1}}G^{}_{{\bf k}{\bf k}^{}_{1}}(\omega)$,
are (in spin space)
\begin{align}
G^{}_{dL}(\omega)&=g^{}_{d}(\omega)[e^{-i\phi^{}_{L}}V^{\dagger}_{Ld}G^{}_{LL}(\omega)+e^{i\phi^{}_{R}}V^{\dagger}_{Rd}G^{}_{RL}(\omega)]\ ,\nonumber\\
G^{}_{Ld}(\omega)&=[e^{i\phi^{}_{L}}G^{}_{LL}(\omega)V^{}_{Ld}+e^{-i\phi^{}_{R}}
G^{}_{LR}(\omega)V^{}_{Rd}]g^{}_{d}(\omega)
\ .
\label{dLLd}
\end{align}
\begin{widetext}
\noindent The integrand in the rate Eq. (\ref{ratem})
 is the lesser Green's function $[
{\rm Int}^{L}_{}(\omega)]^{<}$, where
${\rm Int}^{L}_{}(\omega)$, using Eqs. (\ref{dLLd}), is
\begin{align}
{\rm Int}^{L}_{}(\omega)
&=G^{}_{LR}(\omega){\cal A}^{}_{RL}(\omega)+G^{}_{LL}(\omega)a^{}_{LL}(\omega)-{\cal A}^{}_{LR}(\omega)G^{}_{RL}(\omega)-a^{}_{LL}(\omega)G^{}_{LL}(\omega)\ .
\label{int}
\end{align}
Here we have  introduced the notations
\begin{align}
a^{}_{LL}(\omega)&=V^{}_{Ld}g^{}_{d}(\omega)V^{\dagger}_{Ld}\ ,\ \ \ a^{}_{RR}(\omega)=V^{}_{Rd}g^{}_{d}(\omega)V^{\dagger}_{Rd}\ ,\nonumber\\
{\cal A}^{}_{LR}(\omega)&=V^{}_{LR}+e^{i\Phi}V^{}_{Ld}g^{}_{d}(\omega)V^{\dagger}_{Rd}\ ,\ \ {\cal A}^{}_{RL}(\omega)=V^{\dagger}_{LR}+e^{-i\Phi}V^{}_{Rd}g^{}_{d}(\omega)V^{\dagger}_{Ld}\ .
\label{use}
\end{align}
(In the main text we present ${\cal A}$ in dimensionless units.)
Notice that  $g_{d}(\omega)$ is a real matrix and consequently ${\cal A}^{}_{RL}={\cal A}^{\dagger}_{LR}$.
The Dyson equation for $G_{LL}$ can be written in two equivalent forms
\begin{align}
G^{}_{LL}(\omega)&=g^{}_{L}(\omega)+e^{-i\phi^{}_{L}}G^{}_{Ld}(\omega)V^{\dagger}_{Ld}g^{}_{L}(\omega)+
G^{}_{LR}(\omega)V^{\dagger}_{LR}g^{}_{L}(\omega)\ ,\nonumber\\
&=g^{}_{L}(\omega)+e^{i\phi^{}_{L}}g^{}_{L}(\omega)V^{}_{Ld}G^{}_{dL}(\omega)+g^{}_{L}(\omega)V^{}_{LR}G^{}_{RL}(\omega)\ .
\label{tfo}
\end{align}
Inserting Eqs. (\ref{dLLd}) then gives
\begin{align}
G^{}_{LL}(\omega)&=g^{}_{L}(\omega)+g^{}_{L}(\omega)\Big (a^{}_{LL}(\omega)G^{}_{LL}(\omega)+{\cal A}^{}_{LR}(\omega)G^{}_{RL}(\omega)\Big )\nonumber\\
&=g^{}_{L}(\omega)+\Big (G^{}_{LL}(\omega)a^{}_{LL}(\omega)+G^{}_{LR}(\omega){\cal A}^{}_{RL}(\omega)\Big )g^{}_{L}(\omega)
\ .
\label{LL2}
\end{align}
From Eqs. (\ref{LL2}) we find
\begin{align}
&G^{}_{LL}(\omega)[g^{-1}_{L}(\omega)-a^{}_{LL}(\omega)]=1
+G^{}_{LR}(\omega){\cal A}^{}_{RL}(\omega)\ ,\nonumber\\
&[g^{-1}_{L}(\omega)-a^{}_{LL}(\omega)]G^{}_{LL}(\omega)=1+{\cal A}^{}_{LR}(\omega)G^{}_{RL}(\omega)\ .
\label{GLLn}
\end{align}
Inserting these expressions into Eq. (\ref{int}) yields
\begin{align}
{\rm Int}^{L}_{}(\omega)&=G^{}_{LR}(\omega){\cal A}^{}_{RL}(\omega)
+\Big (1
+G^{}_{LR}(\omega){\cal A}^{}_{RL}(\omega)\Big )
[g^{-1}_{L}(\omega)-a^{}_{LL}(\omega)]^{-1}a^{}_{LL}(\omega)
\nonumber\\
&-{\cal A}^{}_{LR}(\omega)G^{}_{RL}(\omega)-a^{}_{LL}(\omega)
[g^{-1}_{L}(\omega)-a^{}_{LL}(\omega)]^{-1}\Big (
1+{\cal A}^{}_{LR}(\omega)G^{}_{RL}(\omega)\Big )\nonumber\\
&=
G^{}_{LR}(\omega){\cal A}^{}_{RL}(\omega)[1-g^{}_{L}(\omega)a^{}_{LL}(\omega)]^{-1}-[1-a^{}_{LL}(\omega)g^{}_{L}(\omega)]^{-1}{\cal A}^{}_{LR}(\omega)G^{}_{RL}(\omega)\ ,
\label{intn2}
\end{align}
where the final equality results from the fact that
 $g^{}_{L}$ is proportional to the unit matrix.
The Dyson equations for $G_{LR}$ and $G_{RL}$ [Eqs. (\ref{defG})] are
\begin{align}
G^{}_{LR}(\omega)&=[G^{}_{LL}(\omega)V^{}_{LR}+e^{i\phi^{}_{R}}
G^{}_{Ld}V^{\dagger}_{Rd}]g^{}_{R}(\omega)\ , \nonumber\\
G^{}_{RL}(\omega)&=g^{}_{R}(\omega)[e^{-i\phi^{}_{R}}V^{}_{Rd}G^{}_{dL}(\omega)+V^{\dagger}_{LR}G^{}_{LL}(\omega)]\ .
\label{LR}
\end{align}
Inserting
Eqs. (\ref{dLLd}) into Eqs. (\ref{LR}) and then using
Eqs. (\ref{GLLn}), gives
\begin{align}
G^{}_{LR}(\omega)&=[g^{-1}_{L}(\omega)-a^{}_{LL}(\omega)]^{-1}{\cal A}^{}_{LR}(\omega)
)[g^{-1}_{R}(\omega)-a^{}_{RR}(\omega)]^{-1}\nonumber\\
&\times\Big [1-{\cal A}^{}_{RL}(\omega)[g^{-1}_{L}(\omega)-a^{}_{LL}(\omega)]^{-1}{\cal A}^{}_{LR}(\omega)
)[g^{-1}_{R}(\omega)-a^{}_{RR}(\omega)]^{-1}\Big ]^{-1}\ ,\nonumber\\
G^{}_{RL}(\omega)&=\Big [1-[g^{-1}_{R}(\omega)-a^{}_{RR}(\omega)]^{-1}{\cal A}^{}_{RL}(\omega)
)[g^{-1}_{L}(\omega)-a^{}_{LL}(\omega)]^{-1}{\cal A}^{}_{LR}(\omega)\Big ]^{-1}\nonumber\\
&\times[g^{-1}_{R}(\omega)-a^{}_{RR}(\omega)]^{-1}{\cal A}^{}_{RL}(\omega)
)[g^{-1}_{L}(\omega)-a^{}_{LL}(\omega)]^{-1}\ .
\end{align}
Thus, the final expression is
${\rm Int}^{L}_{}(\omega)
=\{\ldots\}g^{-1}_{L}(\omega)-g^{-1}_{L}(\omega)\{\dots\}$,
with
\begin{align}
\{\dots\}&
={\cal G}^{}_{LL}(\omega){\cal A}^{}_{LR}(\omega){\cal G}^{}_{RR}(\omega){\cal A}^{}_{RL}(\omega){\cal G}^{}_{LL}(\omega)
\nonumber\\
&+{\cal G}^{}_{LL}(\omega){\cal A}^{}_{LR}(\omega){\cal G}^{}_{RR}(\omega){\cal A}^{}_{RL}(\omega){\cal G}^{}_{LL}(\omega)
{\cal A}^{}_{LR}(\omega){\cal G}^{}_{RR}(\omega){\cal A}^{}_{RL}(\omega){\cal G}^{}_{LL}(\omega)+\dots\  ,
\label{curly}
\end{align}
where
\begin{align}
{\cal G}^{}_{RR}(\omega)=[g^{-1}_{R}(\omega)-a^{}_{RR}(\omega)]^{-1}_{}\ ,
\end{align}
and similarly for ${\cal G}_{LL}(\omega)$.

The lowest-order expansion in the tunneling of  the terms in Eq. (\ref{curly}) yields
\begin{align}
{\rm Int}^{L}_{}(\omega)
&\approx g^{}_{L}(\omega){\cal A}^{}_{LR}(\omega)
g^{}_{R}(\omega){\cal A}^{}_{RL}(\omega)-{\cal A}^{}_{LR}(\omega)g^{}_{R}(\omega){\cal A}^{}_{RL}(\omega)g^{}_{L}(\omega)\ .
\end{align}
It remains to apply the Langreth rules,  \cite{Langreth} to obtain
\begin{align}
[{\rm Int}^{L}_{}(\omega)]^{<}_{}
&=4\pi^{2}{\cal N}^{}_{L}{\cal N}^{}_{R}{\cal A}^{}_{LR}(\omega){\cal A}^{}_{RL}(\omega)[f^{}_{R}(\omega)-f^{}_{L}(\omega)]\ ,
\label{ratef}
\end{align}
where ${\cal N}^{}_{L(R)}$ is the density of states at the Fermi level in the left (right) lead.

\end{widetext}
\section{Tunneling amplitude }
\label{amplitude}

Our derivation of the tunneling amplitude of an electron  through a potential barrier is based on the one given in Ref. \onlinecite{Shahbazyan}.  These authors considered 
an  electron subjected to the linear Rashba spin-orbit interaction. \cite{Bychkov} In Ref. \onlinecite{com1} we extended their treatment to include the effect of the Zeeman interaction, for the case where the  Fermi energy of the electrons in the leads exceeds the  energy of the potential barrier. Here we consider the opposite situation,  where the energy of the tunneling electron  within the tunneling region is negative. This case was studied in Ref. \onlinecite{Shahbazyan}; we extend that study to include the Zeeman interaction, for a Zeeman field  perpendicular to the tunneling link and to  the  magnetic field induced by the spin-orbit interaction.
Adopting units in which $\hbar=1$, the Hamiltonian of an electron in the tunneling region is
\begin{align}
{\cal H}&=\frac{1}{2m^{\ast}}\Big(-i\frac{d}{d{\bf s}}-\frac{e}{c}{\bf A}\Big)^{2}_{}
\nonumber\\
&+\frac{k^{}_{\rm so}}{m^{\ast}_{}}\hat{\bf n}\cdot\sig\times\Big(-i\frac{d}{d{\bf s}}-\frac{e}{c}{\bf A}\Big)-
{\bf B}\cdot\sig\ ,
\label{ham}
\end{align}
where $\sig=[\sigma^{}_{x},\sigma^{}_{y},\sigma^{}_{z}]$ is the vector of the Pauli matrices, and ${\bf s}$ is the coordinate along the tunneling path (assumed below to be on a straight line 
in the $x-y$ plane). In Eq. (\ref{ham}),   ${\bf A}$ is the vector potential, chosen to be along the direction of ${\bf s}$, 
$m^{\ast}$ is the (effective) mass, $\hat{\bf n}$ is the direction of the electric field creating the
spin-orbit interaction, whose strength is $k_{\rm so}$ in momentum units, and ${\bf B}$
is the external magnetic field (in energy units), which  is  along the
$\hat{\bf z}-$direction.

For a
plane-wave solution with a wave vector ${\bf k}$
directed along ${\bf s}$,
 the  magnetic field induced by the spin-orbit interaction is
\begin{align}
{\bf B}^{}_{\rm so}({\bf k})=\frac{kk^{}_{\rm so}}{m^{\ast}_{}}\hat{\bf e}\ ,\ \ {\rm with} \ \ \ \hat{\bf e}=\hat{\bf n}\times\hat{\bf k}\ .
\label{BQ}
\end{align}
This effective magnetic field depends on the direction of ${\bf k}$, that is, on the direction of ${\bf s}$.
For $\hat{\bf n}$ along the $\hat{\bf z}-$direction,
${\bf B}\perp{\bf B}^{}_{\rm so}({\bf k})$ and $\hat{\bf e}$ is a unit vector in the $x-y$ plane. 
The vector potential
${\bf A}$, which represents the orbital effect of the magnetic field,  can be gauged out from the
Hamiltonian, to reappear as the  Aharonov-Bohm phase factor multiplying
the tunneling amplitude (see  the main text); hence it is  ignored in the following.
The spin-dependent propagator (i.e., the retarded Green's function) is a 2$\times$2 matrix
in spin space.  When the energy of the tunneling electron is negative (i.e., the Fermi energy in the leads is smaller than the potential barrier representing the tunneling region)
then
 $E=-1/(2m^{\ast}a^{2})$, where $a$ measures the extent of the localized wave function.   \cite{Shahbazyan}
 The tunneling amplitude, i.e., the propagator $G({\bf s})$, is
\begin{align}
&G({\bf s})=\int dk e^{iks}[-1/(2m^{\ast}a^{2})+i0^{+}_{}-{\cal H}({\bf k})]^{-1}\ ,
\label{1Gs}
\end{align}
where
\begin{align}
{\cal H}({\bf k})=\frac{k^{2}}{2m^{\ast}}-\frac{kk^{}_{\rm so}}{m^{\ast}}\hat{\bf e}\cdot\sig-B\sigma^{}_{z}\ .
\end{align}
The integral in Eq. (\ref{1Gs})  is evaluated
by the Cauchy theorem, for the case where  the coordinate $\hat{\bf s}=s\hat{\bf k}$ is along a straight line, assuming that $s>0$,
\begin{align}
&2m^{\ast}
\int dk e^{iks}\frac{-(1/a^{2})-k^{2}-2m^{\ast}B\sigma^{}_{z}-2kk^{}_{\rm so}\hat{\bf e}\cdot\sig
}{[(1/a^{2})+k^{2}]^{2}-(2kk^{}_{\rm so})^{2}-(2m^{\ast}B)^{2}}\nonumber\\
&=-\frac{i\pi m^{\ast}}
{k^{2}_{+}-k^{2}_{-}}\Big (\frac{e^{ik^{}_{+}s}}{k^{}_{+}}[\frac{1}{a^{2}}+k^{2}_{+}+2m^{\ast}B\sigma^{}_{z}+
2k^{}_{+}k^{}_{\rm so}\hat{\bf e}\cdot\sig]
\nonumber\\
&-
\frac{e^{ik^{}_{-}s}}{k^{}_{-}}[\frac{1}{a^{2}}+k^{2}_{-}+2m^{\ast}B\sigma^{}_{z}+
2k^{}_{-}k^{}_{\rm so}\hat{\bf e}\cdot\sig]\Big )\ .
\label{prop}
\end{align}
Here, $k_{+}$ and $k_{-}$ are the roots of the denominator,
\begin{align}
&[(1/a^{2})+k^{2}]^{2}-(2kk^{}_{\rm so})^{2}-(2m^{\ast}B)^{2}\nonumber\\
&=(k^{2}_{+}-k^{2}_{})(k^{2}_{-}
-k^{2}_{})\ ,
\label{kpm}
\end{align}
where
\begin{align}
k^{2}_{\pm}=-\frac{1}{a^{2}}+2k^{2}_{\rm so}\pm 2\sqrt{-\frac{k^{2}_{\rm so}}{a^{2}}+k^{4}_{\rm so}+(m^{\ast}B)^{2}}\ .
\end{align}

Focusing on the case $2m^{\ast}Ba^{4}< 1$, we introduce the variables $k_{1}$ and $k_{2}$
\begin{align}
k_{\pm}=ik^{}_{1}\pm k^{}_{2}\ ,
\end{align}
which obey
\begin{align}
k^{2}_{2}-k^{2}_{1}&=-\frac{1}{a^{2}}+2k^{2}_{\rm so}\ ,\nonumber\\
ik^{}_{1}k^{}_{2}&=\sqrt{-\frac{k^{2}_{\rm so}}{a^{2}}+k^{4}_{\rm so}+(m^{\ast}B)^{2}}\ ,
\end{align}
and therefore
\begin{align}
k^{}_{1}&=\Big [\frac{1}{2a^{2}}-k^{2}_{\rm so}+\sqrt{\frac{1}{4a^{4}}-(m^{\ast}B)^{2}}\Big ]^{1/2}_{}\ ,\nonumber\\
k^{}_{2}&=\Big [-\frac{1}{2a^{2}}+k^{2}_{\rm so}+\sqrt{\frac{1}{4a^{4}}-(m^{\ast}B)^{2}}\Big ]^{1/2}_{}\ .
\end{align}
In terms of these variables, the scalar part of the expression on the right hand-side of Eq. (\ref{prop}) is
\begin{align}
\frac{\pi m^{\ast}e^{-k^{}_{1}s}}{k^{}_1k^{}_2(k_1^2+k_2^2)}&\Big[k_2^{}\big(-k_1^2-k^{2}_{\rm so}\big)\cos(k^{}_{2}s)\nonumber\\
&+k^{}_1\big(k_2^2-k^{2}_{\rm so}\big)\sin(k^{}_{2}s)
\big]\ ,
\label{sca}
\end{align}
the $\sigma^{}_{z}$ part is
\begin{align}
&-\frac{\pi m^{\ast}e^{-k^{}_{1}s}}{k^{}_1k^{}_2(k_1^2+k_2^2)}\big[k^{}_{2}\cos(k^{}_{2}s)+k^{}_{1}\sin(k^{}_{2}s)
\big]m^\ast B\sigma^{}_z
\ ,
\label{sigmaz}
\end{align}
and the part related to $\hat{\bf e}\cdot\sig$ is
\begin{align}
&-\frac{\pi m^{\ast}e^{-k^{}_{1}s}}{k^{}_1k^{}_2}
[ik^{}_{\rm so}\sin(k^{}_{2}s)]\hat{\bf e}\cdot\sig\ .
\label{edots}
\end{align}
Adopting the  plausible assumption that $[m^{\ast}a^{2}]^{-1}$ is larger than the spin-orbit and the Zeeman energies,
\begin{align}
m^{\ast}Ba^{2}\ll1\ ,\ \ (k^{}_{\rm so}a)^{2}\ll 1
\label{ineq}
\end{align}
we find that
\begin{align}
k^{}_{1}\approx1/a\ ,\ \ k^{}_{2}\approx\sqrt{k^{2}_{\rm so}-(m^{\ast}Ba)^{2}}\ .
\label{k1k2}
\end{align}
The propagator is then
\begin{align}
&G({\bf s})=-\pi m^{\ast}ae^{-as}\nonumber\\
&\times\Big (\cos(k^{}_{2}s)+\frac{\sin(k^{}_{2}s)}{k^{}_{2}}[ik^{}_{\rm so}\hat{\bf e}\cdot\sig +m^{\ast}Ba\sigma^{}_{z}]\Big )\ .
\label{propsoB}
\end{align}
In the main text we replace the prefactor of $G(s)$ by $J$, ignoring for simplicity its dependence on the length of the bond, i.e.
 $J=-\pi m^{\ast}a\exp[-as]$.\cite{comm}

In the  limit of zero   Zeeman field ${\bf B}$,  Eq. (\ref{propsoB}) for the propagator reduces to
\begin{align}
G(s)&=-\pi m^{\ast}a e^{- s/a}
\Big (\cos(k^{}_{\rm so}s)
+i\sin(k^{}_{\rm so}s)\hat{\bf e}\cdot\sig\Big )\nonumber\\
&=-\pi m^{\ast}a e^{- s/a}\exp[ik^{}_{\rm so}s\hat{\bf e}\cdot\sig]\ .
\label{prop1}
\end{align}
This unitary form is the one
expected when only the spin-orbit interaction is active. \cite{Meir}
On the other hand, when there is only the Zeeman field and  $k_{2}$ is purely imaginary,  the propagator is
\begin{align}
G(s)&=-\pi m^{\ast}a e^{-s/a}[\cosh(\kappa s)+\sinh(\kappa s)\sigma^{}_{z}]\ ,
\end{align}
where
\begin{align}
\kappa
\approx m^{\ast}_{}Ba\ .
\label{kappa}
\end{align}
[Note that $\kappa a\ll1$, see Eqs. (\ref{ineq}).]
In this case  the oscillations as a function of the distance $s$, which are essentially due to the spin-orbit interaction, are absent. In fact, they disappear once
$m^{\ast}(Ba)^{2}$ (the Zeeman energy divided by the energy at the barrier,
$[m^{\ast}a^{2}]^{-1}$) exceeds the spin-orbit energy.


\section{The interference terms in the transmission of a triangular interferometer }
\label{triangle}

Here we consider  the interference contributions to the transmission matrices,  ${\cal T}^{\rm in}_{L}$ and ${\cal T}^{\rm in}_{R}$ [Eqs. (\ref{TLR}) and (\ref{TRL}), respectively] in the presence of both the Rashba and the Zeeman interactions. In this case,  the tunneling amplitude of each bond is given by Eq. (\ref{propsoB}), and is not proportional to a unitary matrix as in  the presence of the spin-orbit interaction alone;  the Zeeman interaction also changes  the Green's function of the decoupled dot,  $g_{d}(\omega)$, into a matrix, since by Eq. (\ref{gd})
\begin{align}
g^{}_{d}(\omega)=[\omega-\epsilon^{}_{0}-\sigma^{}_{z}B]/[(\omega-\epsilon^{}_{0})^{2}-B^{2}]\ .
\label{gdZ}
\end{align}
Below, we first ignore the Zeeman interaction on the dot  [i.e., we assume that $g_{d}(\omega)=(\omega-\epsilon^{}_{0})^{-1}$] and consider only
the product
$V^{}_{LR}V^{}_{Rd}V^{}_{dL}$ [see e.g., Eq. (\ref{TLR})]; we then  investigate this product when the  term $\sigma^{}_{z}B$ in Eq. (\ref{gdZ}) is accounted for.

To consider the product
$V^{}_{LR}V^{}_{Rd}V^{}_{dL}$ for an arbitrary triangular interferometer, it is convenient
to introduce the shorthand notations
\begin{align}
t^{}_{ij}=\tan(k^{}_{2}d^{}_{ij})\ ,\  \overline{k}^{}_{\rm so}=k^{}_{\rm so}/k^{}_{2}\ ,\  \overline{B}=m^{\ast}Ba/k^{}_{2}
\ .
\end{align}
In the absence of the Zeeman field $k_{2}^{}=k^{}_{\rm so}$ and $\overline{k}^{}_{\rm so}=1$,  while when there is no spin-orbit interaction $k_{2}=im^{\ast}Ba$,  $t_{ij}$ is purely imaginary, and $\overline{B}=-i$ [see Eqs. (\ref{k1k2})].
With these notations,
\begin{align}
&\frac{V^{}_{LR}V^{}_{Rd}V^{}_{dL}/J^{3}}{\cos(k^{}_{2}d^{}_{LR})
\cos(k^{}_{2}d^{}_{Rd})\cos(k^{}_{2}d^{}_{dL})}\nonumber\\
&=(1+X^{}_{LR})(1+X^{}_{Rd})(1+X^{}_{dL})\ ,
\label{VVV}
\end{align}
where
\begin{align}
X^{}_{ij}=t^{}_{ij}[i\overline{k}^{}_{\rm so}\hat{\bf e}^{}_{ij}\cdot\sig
+\overline{B}\sigma^{}_{z}]\ .
\label{AX}
\end{align}
\begin{widetext}
Explicitly,
\begin{align}
X^{}_{LR}+X^{}_{Rd}+X^{}_{dL}=\overline{B}\sigma^{}_{z}[t^{}_{LR}+t^{}_{Rd}+t^{}_{dL}]+i\overline{k}^{}_{\rm so}\sig\cdot[
t^{}_{LR}\hat{\bf e}^{}_{LR}+t^{}_{Rd}\hat{\bf e}^{}_{Rd}+t^{}_{dL}
\hat{\bf e}^{}_{dL}]\ ,
\label{LA}
\end{align}
\begin{align}
X^{}_{LR}X^{}_{Rd}+&X^{}_{Rd}X^{}_{dL}+X^{}_{LR}X^{}_{dL}=t^{}_{LR}t^{}_{Rd}
\Big (-\overline{k}^{2}_{\rm so}\hat{\bf e}^{}_{LR}\cdot\hat{\bf e}^{}_{Rd}+\overline{B}^{2}
-i\overline{k}^{2}_{\rm so}\hat{\bf e}^{}_{LR}\times\hat{\bf e}^{}_{Rd}\cdot\sig-\overline{k}^{}_{\rm so}\overline{B}[\hat{\bf e}^{}_{LR}\times\hat{\bf z}+\hat{\bf z}\times\hat{\bf e}^{}_{Rd}]\cdot\sig\Big )
\nonumber\\
&+t^{}_{Rd}t^{}_{dL}
\Big (-\overline{k}^{2}_{\rm so}\hat{\bf e}^{}_{Rd}\cdot\hat{\bf e}^{}_{dL}+\overline{B}^{2}
-i\overline{k}^{2}_{\rm so}\hat{\bf e}^{}_{Rd}\times\hat{\bf e}^{}_{dL}\cdot\sig-\overline{k}^{}_{\rm so}\overline{B}[\hat{\bf e}^{}_{Rd}\times\hat{\bf z}+\hat{\bf z}\times\hat{\bf e}^{}_{dL}]\cdot\sig\Big )
\nonumber\\
&+t^{}_{LR}t^{}_{dL}
\Big (-\overline{k}^{2}_{\rm so}\hat{\bf e}^{}_{LR}\cdot\hat{\bf e}^{}_{dL}+\overline{B}^{2}
-i\overline{k}^{2}_{\rm so}\hat{\bf e}^{}_{LR}\times\hat{\bf e}^{}_{dL}\cdot\sig-\overline{k}^{}_{\rm so}\overline{B}[\hat{\bf e}^{}_{LR}\times\hat{\bf z}+\hat{\bf z}\times\hat{\bf e}^{}_{dL}]\cdot\sig\Big )\ ,
\label{AA}
\end{align}
and
\begin{align}
&X^{}_{LR}X^{}_{Rd}X^{}_{dL}=t^{}_{LR}t^{}_{Rd}t^{}_{dL}
 (-\overline{k}^{2}_{\rm so}\hat{\bf e}^{}_{LR}\cdot\hat{\bf e}^{}_{Rd}+\overline{B}^{2}
-i\overline{k}^{2}_{\rm so}\hat{\bf e}^{}_{LR}\times\hat{\bf e}^{}_{Rd}\cdot\sig-\overline{k}^{}_{\rm so}\overline{B}[\hat{\bf e}^{}_{LR}\times\hat{\bf z}+\hat{\bf z}\times\hat{\bf e}^{}_{Rd}]\cdot\sig )\nonumber\\
&\hspace{5cm}\times[i\overline{k}^{}_{\rm so}\hat{\bf e}^{}_{dL}\cdot\sig
+\overline{B}\sigma^{}_{z}]\nonumber\\
&=t^{}_{LR}t^{}_{Rd}t^{}_{dL}
\Big (-i\overline{k}^{2}_{\rm so}\overline{B}[\hat{\bf e}^{}_{LR}\times\hat{\bf e}^{}_{Rd}\cdot\hat{\bf z}+\hat{\bf e}^{}_{dL}\times\hat{\bf e}^{}_{LR}\cdot\hat{\bf z}+
\hat{\bf e}^{}_{Rd}\times\hat{\bf e}^{}_{dL}\cdot\hat{\bf z}]\nonumber\\
&-i\overline{k}^{3}_{\rm so}\sig\cdot[(\hat{\bf e}^{}_{LR}\cdot\hat{\bf e}^{}_{Rd})\hat{\bf e}^{}_{dL}+(\hat{\bf e}^{}_{dL}\cdot\hat{\bf e}^{}_{Rd})\hat{\bf e}^{}_{LR}-
(\hat{\bf e}^{}_{dL}\cdot\hat{\bf e}^{}_{LR})\hat{\bf e}^{}_{Rd}]+\overline{B}^{3}\sigma^{}_{z}+i\overline{k}^{}_{\rm so}\overline{B}^{2}_{}\sig\cdot[\hat{\bf e}^{}_{dL}+\hat{\bf e}^{}_{LR}-\hat{\bf e}^{}_{Rd}]\nonumber\\
&+\overline{k}^{2}_{\rm so}\overline{B}\sigma^{}_{z}[-\hat{\bf e}^{}_{LR}\cdot\hat{\bf e}^{}_{Rd}+\hat{\bf e}^{}_{dL}\cdot\hat{\bf e}^{}_{LR}-\hat{\bf e}^{}_{Rd}\cdot\hat{\bf e}^{}_{dL}]\Big )\ .
\label{AAA}
\end{align}

In the absence of the spin-orbit interaction,
\begin{align}
&\frac{V^{}_{LR}V^{}_{Rd}V^{}_{dL}/J^{3}}{\cosh(\kappa d^{}_{LR})
\cosh(\kappa d^{}_{Rd})\cosh(\kappa d^{}_{dL})}\Big |^{}_{k^{}_{\rm so}=0}=
1+
\tau^{}_{LR}\tau^{}_{Rd}+\tau ^{}_{Rd}\tau ^{}_{dL}+\tau ^{}_{dL}\tau ^{}_{LR}+
\sigma^{}_{z}\Big (\tau^{}_{LR}+\tau^{}_{Rd}
+\tau ^{}_{dL}+
\tau ^{}_{LR}\tau^{}_{Rd}\tau^{}_{dL}\Big )\ ,
\label{noSO}
\end{align}
where $\kappa$ is given in Eq. (\ref{kappa}) and $\tau_{ij}=\tanh(\kappa d^{}_{ij})$.
In this case $V^{}_{LR}V^{}_{Rd}V^{}_{dL}=V^{}_{RL}V^{}_{Ld}V^{}_{dR}$ is a real matrix, invariant under $L\leftrightarrow R$; as a result,  the charge conductance and the magnetization rates are proportional to $\cos(\Phi)$,, and they are both conserved. In the absence of   the spin-orbit interaction, the magnetization rates are solely along the $\hat{\bf z}-$axis, that is, along the direction of the Zeeman field.

In the absence of the Zeeman interaction
\begin{align}
&\frac{V^{}_{LR}V^{}_{Rd}V^{}_{dL}/J^{3}}{\cos(k^{}_{2}d^{}_{LR})
\cos(k^{}_{2}d^{}_{Rd})\cos(k^{}_{2}d^{}_{dL})}\Big |^{}_{B=0}=1-[t^{}_{LR}t^{}_{Rd}\hat{\bf e}^{}_{LR}\cdot\hat{\bf e}^{}_{Rd}+t^{}_{LR}t^{}_{dL}\hat{\bf e}^{}_{LR}\cdot\hat{\bf e}^{}_{dL}+
t^{}_{dL}t^{}_{Rd}\hat{\bf e}^{}_{dL}\cdot\hat{\bf e}^{}_{Rd}]\nonumber\\
&+i\sig\cdot\Big [t^{}_{dL}\hat{\bf e}^{}_{dL}+t^{}_{Rd}\hat{\bf e}^{}_{Rd}+t^{}_{LR}\hat{\bf e}^{}_{LR}-[t^{}_{LR}t^{}_{dL}(\hat{\bf e}^{}_{LR}\times\hat{\bf e}^{}_{dL})+t^{}_{LR}t^{}_{Rd}(\hat{\bf e}^{}_{LR}\times\hat{\bf e}^{}_{Rd})+t^{}_{Rd}t^{}_{dL}(\hat{\bf e}^{}_{Rd}\times\hat{\bf e}^{}_{dL})]\nonumber\\
&-t^{}_{LR}t^{}_{Rd}t^{}_{dL}[(\hat{\bf e}^{}_{LR}\cdot\hat{\bf e}^{}_{Rd})\hat{\bf e}^{}_{dL}
+(\hat{\bf e}^{}_{dL}\cdot\hat{\bf e}^{}_{Rd})\hat{\bf e}^{}_{LR}
-(\hat{\bf e}^{}_{dL}\cdot\hat{\bf e}^{}_{LR})\hat{\bf e}^{}_{Rd}]\Big ]\ .
\label{noZ}
\end{align}
This case is discussed in great detail in Sec. \ref{trnoB}. Equating Eq. (\ref{noZ}) to Eq. (\ref{p1}) yields
\begin{align}
&\frac{{\bf c}^{}_{L}}{\cos(k^{}_{2}d^{}_{LR})
\cos(k^{}_{2}d^{}_{Rd})\cos(k^{}_{2}d^{}_{dL})}=-[t^{}_{LR}t^{}_{dL}(\hat{\bf e}^{}_{LR}\times\hat{\bf e}^{}_{dL})+t^{}_{LR}t^{}_{Rd}(\hat{\bf e}^{}_{LR}\times\hat{\bf e}^{}_{Rd})+t^{}_{Rd}t^{}_{dL}(\hat{\bf e}^{}_{Rd}\times\hat{\bf e}^{}_{dL})]\nonumber\\
&+t^{}_{dL}\hat{\bf e}^{}_{dL}+t^{}_{Rd}\hat{\bf e}^{}_{Rd}+t^{}_{LR}\hat{\bf e}^{}_{LR}
-t^{}_{LR}t^{}_{Rd}t^{}_{dL}[(\hat{\bf e}^{}_{LR}\cdot\hat{\bf e}^{}_{Rd})\hat{\bf e}^{}_{dL}
+(\hat{\bf e}^{}_{dL}\cdot\hat{\bf e}^{}_{Rd})\hat{\bf e}^{}_{LR}
-(\hat{\bf e}^{}_{dL}\cdot\hat{\bf e}^{}_{LR})\hat{\bf e}^{}_{Rd}]\Big )\ .
\end{align}
Projecting ${\bf c}^{}_L$ on $\hat{\bf e}^{}_{LR}$  confirms  Eq. (\ref{MMM1}). Projections on the transverse directions does not yield simple relations between the transverse magnetization rates of the two reservoirs. An exception is the
equilateral triangle, for which
\begin{align}
{\bf  c}^{}_L=-\sin^2(k^{}_2d)[\cos(k^{}_2d)(\sqrt{3}/2)\hat{\bf z}+\sin(k^{}_2d)\hat{\bf e}^{}_{Rd}]\ ,
\end{align}
and therefore
\begin{align}
{\bf  c}^{}_R=-\sin^2(k^{}_2d)[\cos(k^{}_2d)(\sqrt{3}/2)\hat{\bf z}+\sin(k^{}_2d)\hat{\bf e}^{}_{Ld}]\ .
\end{align}
In this special case we find that $c^{}_{Rz}=c^{}_{Lz}$, hence $\chi^{}_R=\chi^{}_L$. Setting $\hat{\bf d}^{}_{LR}=\hat{\bf x}$ yields $\varphi^{}_R=-\varphi^{}_L$.

Returning to the case where both  spin-orbit and  Zeeman interactions are present in the tunneling amplitude, we first ignore the effect of the Zeeman interaction on the dot and examine the scalar part of the product
$V^{}_{LR}V^{}_{Rd}V^{}_{dL}$ (which determines the charge conductance), that is
\begin{align}
&{\rm Tr}\Big \{
\frac{V^{}_{LR}V^{}_{Rd}V^{}_{dL}/J^{3}}{
\cos(k^{}_{2}d^{}_{LR})
\cos(k^{}_{2}d^{}_{Rd})
\cos(k^{}_{2}d^{}_{dL})}\Big\}
=2\Big (1+t^{}_{LR}t^{}_{Rd}
 [-\overline{k}^{2}_{\rm so}\hat{\bf e}^{}_{LR}\cdot\hat{\bf e}^{}_{Rd}+\overline{B}^{2}]
+t^{}_{Rd}t^{}_{dL}
[-\overline{k}^{2}_{\rm so}\hat{\bf e}^{}_{Rd}\cdot\hat{\bf e}^{}_{dL}+\overline{B}^{2}]
\nonumber\\
&+t^{}_{LR}t^{}_{dL}
[-\overline{k}^{2}_{\rm so}\hat{\bf e}^{}_{LR}\cdot\hat{\bf e}^{}_{dL}+\overline{B}^{2}]-i\overline{k}^{2}_{\rm so}\overline{B}
t^{}_{LR}t^{}_{Rd}t^{}_{dL}
[\hat{\bf e}^{}_{LR}\times\hat{\bf e}^{}_{Rd}\cdot\hat{\bf z}+\hat{\bf e}^{}_{dL}\times\hat{\bf e}^{}_{LR}\cdot\hat{\bf z}+
\hat{\bf e}^{}_{Rd}\times\hat{\bf e}^{}_{dL}\cdot\hat{\bf z}]\Big )\ .
\label{triple}
\end{align}
Here we find a dramatic difference as compared to the case where the Zeeman interaction is absent: the appearance of the last
term on the right hand-side of Eq. (\ref{triple}). As shown in Sec. \ref{trB}, this term modifies  the Aharonov-Bohm oscillations of the magnetoconductance  as a function of the  flux $\Phi$, which now acquire an additional dependence upon $\sin(\Phi)$. The amplitude of the latter is proportional to the Zeeman field (and thus the result is compatible with the Onsager relations) and vanishes in the absence of the spin-orbit coupling.

Adding the Zeeman energy to the Green's function of the decoupled dot [see Eq. (\ref{gdZ})]
and expanding to linear order in $B$,    yields
\begin{align}
V^{}_{LR}V^{}_{Rd}g^{}_{d}(\omega)V^{}_{dL}\approx \big[V^{}_{LR}V^{}_{Rd}V^{}_{dL}-BV^{}_{LR}V^{}_{Rd}\sigma^{}_zV^{}_{dL}/(\omega-\epsilon^{}_0)\big]/(\omega-\epsilon^{}_0)\ ,
\label{dotZ}
\end{align}
where the tunneling amplitudes in the last term on the right hand-side of Eq. (\ref{dotZ}) should be considered for $B=0$, i.e., they are given by Eqs. (\ref{e}).
Then, noting that
$\sigma^{}_z(\hat{\bf e}^{}_{dL}\cdot\sig)\sigma^{}_z=(\hat{\bf e}^{}_{Ld}\cdot\sig)$, leads to
\begin{align}
V^{}_{LR}V^{}_{Rd}\sigma^{}_zV^{}_{dL}=V^{}_{LR}V^{}_{Rd}V^{}_{Ld}\sigma^{}_z\ ,
\label{dotZ1}
\end{align}
which implies that we need the product
$V^{}_{LR}V^{}_{Rd}V^{}_{Ld}$. The latter is given in
Eq. (\ref{noZ}) with $\hat{\bf e}^{}_{dL}$ replaced by $\hat{\bf e}^{}_{Ld}$.
 The resulting contribution from the Zeeman energy in the Green's function of the dot to the scalar part of Eq. (\ref{dotZ})  then comes from the terms of the form $it^{}_{ij}t^{}_{i'j'}[\hat{\bf e}^{}_{ij}\times\hat{\bf e}^{}_{i'j'}]\cdot\sig$ in Eq. (\ref{noZ}). Multiplying such a term by $\sigma^{}_z$ contributes to the scalar part an imaginary term $it^{}_{ij}t^{}_{i'j'}[\hat{\bf e}^{}_{ij}\times\hat{\bf e}^{}_{i'j'}]\cdot\hat{\bf z}$, which is similar to the contribution of the Zeeman interaction in the tunneling amplitudes  [i.e., the last term in Eq. (\ref{triple})], except that its coefficient now contains $B/(\omega-\epsilon^{}_0)$ instead of $\overline{B}t^{}_{i"j"}$. These  add to the $B\sin(\Phi)$ term  in the charge conductance, in the same way as the last term Eq. (\ref{triple}). When the dot is far from resonance, these contributions are smaller (note that our analysis is not strictly valid when the dot is close to resonance).

 Turning now to the terms yielding the magnetization rates, we focus on those linear in the Zeeman field.
There are two types of such terms. First, there are those linear in $\overline{B}$
 that are included in
 Eqs. (\ref{VVV})-(\ref{AAA}). These should be multiplied by $\sig$ [see, e.g., Eq. (\ref{ML1})]
 and then traced over.
 An examination of Eqs. (\ref
 {VVV})-(\ref{AAA}) shows that they yield
 \begin{align}
&\overline{B} {\rm Tr}\Big \{\Big [
\sigma^{}_{z}[t^{}_{LR}+t^{}_{Rd}+t^{}_{dL}]
 +t^{}_{LR}t^{}_{Rd}t^{}_{dL}\overline{k}^{2}_{\rm so}\sigma^{}_{z}[-\hat{\bf e}^{}_{LR}\cdot\hat{\bf e}^{}_{Rd}+\hat{\bf e}^{}_{dL}\cdot\hat{\bf e}^{}_{LR}-\hat{\bf e}^{}_{Rd}\cdot\hat{\bf e}^{}_{dL}]
 \nonumber\\
&-\overline{k}^{}_{\rm so}\Big (
t^{}_{LR}t^{}_{Rd}
[\hat{\bf e}^{}_{LR}\times\hat{\bf z}+\hat{\bf z}\times\hat{\bf e}^{}_{Rd}]
+t^{}_{Rd}t^{}_{dL}
[\hat{\bf e}^{}_{Rd}\times\hat{\bf z}+\hat{\bf z}\times\hat{\bf e}^{}_{dL}]
+t^{}_{LR}t^{}_{dL}
[\hat{\bf e}^{}_{LR}\times\hat{\bf z}+\hat{\bf z}\times\hat{\bf e}^{}_{dL}]\Big )\cdot\sig
\Big ]\sig\cdot\hat{\el}
 \Big\}\ ,
 \label{ft}
 \end{align}
for the magnetization rate along an arbitrary unit vector $\hat{\el}$.
As seen, the first two terms on the right hand-side of Eq.  (\ref{ft}) contribute only to the magnetization rates along $\hat{\bf z}$, and that contribution is real. As a result, they give rise to a $B\cos(\Phi)$ dependence in $\dot{\bf M}^{L(R)}_{z}$. The last expression on the right hand-side of Eq. (\ref{ft}) necessitates the product $\sig^{}_{\perp}\sig\cdot\hat{\el}$, where the vector $\sig^{}_{\perp}$, which denotes the circular brackets, lies in the $x-y$ plane. Therefore,  terms of this type contribute only to the  in-plane magnetization rates,
 $\dot{\bf M}^{L(R)}_{\perp}$,
and that contribution which is real,  also leads  to a $B\cos(\Phi)$ dependence of the magnetization rates.

The second type of contributions comes from the Zeeman interaction in the Green's function on the dot.
Exploiting Eqs. (\ref{dotZ}) and (\ref{dotZ1}) and the definition (\ref{ML1}) implies
 that the expression for the magnetization rates along the unit vector $\hat{\el}$ due to those comprises terms of the form
\begin{align}
\frac{B}{\omega-\epsilon^{}_{0}}{\rm Tr}\Big
\{
V^{}_{LR}V^{}_{Rd}\sigma^{}_{z}V^{}_{dL}\sig\cdot\hat{\el}\Big \}=\frac{B}{\omega-\epsilon^{}_{0}}{\rm Tr}\Big
\{
V^{}_{LR}V^{}_{Rd}V^{}_{Ld}\sigma^{}_{z}\sig\cdot\hat{\el}\Big \}\ .
\end{align}
As
$V^{}_{LR}V^{}_{Rd}V^{}_{Ld}=U+i{\bf W}\cdot \sig$ at $\overline{B}=0$, $V^{}_{LR}V^{}_{Rd}V^{}_{Ld}\sigma^{}_z=U\sigma^{}_z+i{\bf W}\cdot\hat{\bf z}-[{\bf W}\times\hat{\bf z}]\cdot\sig$.
 From Eq. (\ref{noZ}), $U$ and ${\bf W}$ are real, and therefore the contributions of the Zeeman interaction  to the magnetization rates are  real, generating terms proportional to $\cos(\Phi)$. 
 Thus,  all  magnetization rates due to the Zeeman interaction,  of order $\overline{B}$ or $B/(\omega-\epsilon^{}_{0})$, have coefficients which are even functions of $\Phi$. This is not surprising: the magnetization rates are expected to be odd functions of the magnetic fields. Note that the magnetization rates due to the spin-orbit interaction alone, Eq. (\ref{TLRS}), all have  a $\sin(\Phi)$ dependence.

\end{widetext}



\begin{thebibliography}{99}
\bibitem{comm} As seen in Appendix \ref{triangle}, the $\Phi-$dependent terms are all proportional to the product of the three $J$'s.
\bibitem{comim}
Note that $k_{2}=\sqrt{k^{2}_{\rm so}-(m^{\ast}aB)^{2}}$ always appears in  the terms of Eq. (\ref{triple})
in conjunction with $\tan(k^{}_{2}d^{}_{ij})$, and hence all terms in Eq. (\ref{triple}) are real except for those that are explicitly multiplied by $i$.
\end{thebibliography}
\end{document}